\documentclass[10pt,conference,letterpaper]{IEEEtran}
\usepackage[USenglish]{babel}
\usepackage{times,amsmath}
\usepackage[pdftex]{graphicx}
   \graphicspath{{./pdf/}{./png/}{./jpeg/}{./eps/}}
   \DeclareGraphicsExtensions{.pdf,.png,.jpeg,.eps}
\usepackage{cite}
\usepackage{url}
\usepackage{fmtcount}
\usepackage{color}
\usepackage[tight,footnotesize]{subfigure}
\usepackage{multicol}
\usepackage{amssymb}
\usepackage{wasysym}
\usepackage{dblfloatfix} 
\usepackage{calc}
\usepackage{fix-cm}
\usepackage[activate={true,nocompatibility},final,tracking=true,factor=500,stretch=40,shrink=40]{microtype}
\def\cal{\fam2}
\renewcommand{\d}{{\mathrm d}}

\newcommand{\sgn}{{\mathrm{sgn}}}

\newcommand{\BalphaPM}{{{\cal{B}}^{\boldmath \alpha_+}_{\boldmath \alpha_-}}}
\definecolor{DarkRed}{rgb}{0.5,0,0}
\definecolor{DarkGreen}{rgb}{0,0.5,0}
\definecolor{DarkerGreen}{rgb}{0,0.3333,0}
\definecolor{DarkBlue}{rgb}{0,0,0.75}
\definecolor{RoyalBlue}{rgb}{0,0.1373,0.4000}
\definecolor{NavyBlue}{rgb}{0,0,0.5020}
\definecolor{CobaltBlue}{rgb}{0,0.2784,0.6706}
\definecolor{lightlightgray}{rgb}{0.96875,0.96875,0.96875}
\definecolor{cyan}{rgb}{0,1,1}
\newcommand{\beginlabel}[2]{%
\begin{#1}\label{#2}}
\begin{document}
\pagestyle{plain}
\title{Hidden outlier noise and its mitigation}
\author{\IEEEauthorblockN{Alexei V. Nikitin}
\IEEEauthorblockA{
Nonlinear LLC\\
Wamego, Kansas, USA\\
E-mail: avn@nonlinearcorp.com}
\and
\IEEEauthorblockN{Ruslan L. Davidchack}
\IEEEauthorblockA{Dept, of Mathematics, U. of Leicester\\
Leicester, UK\\
E-mail: rld8@leicester.ac.uk}}
\maketitle
\begin{abstract}
In addition to ever-present thermal noise, various communication and sensor systems can contain significant amounts of interference with outlier (e.g. impulsive) characteristics. Such outlier noise can be efficiently mitigated in real-time using intermittently nonlinear filters. Depending on the noise nature and composition, improvements in the quality of the signal of interest will vary from ``no harm" to substantial. In this paper, we explain in detail why the underlying outlier nature of interference often remains obscured, discussing the many challenges and misconceptions associated with state-of-art analog and/or digital nonlinear mitigation techniques, especially when addressing complex practical interference scenarios. We then focus on the methodology and tools for real-time outlier noise mitigation, demonstrating how the ``excess band" observation of outlier noise enables its efficient in-band mitigation. We introduce the basic real-time nonlinear components that are used for outlier noise filtering, and provide examples of their implementation. We further describe complementary nonlinear filtering arrangements for wide- and narrow-band outlier noise reduction, providing several illustrations of their performance and the effect on channel capacity. Finally, we outline ``effectively analog" digital implementations of these filtering structures, discuss their broader applications, and comment on the ongoing development of the platform for their demonstration and testing.
\end{abstract}
\begin{IEEEkeywords}
Analog filter, digital filter, electromagnetic interference (EMI), impulsive noise, intermittently nonlinear filter, man-made interference, non-Gaussian noise, nonlinear signal processing, outlier noise, technogenic interference.
\end{IEEEkeywords}
\maketitle
\section{Introduction} \label{sec:introduction}
At any given frequency, a linear filter affects all signals proportionally. Therefore, when linear filtering is used to suppress interference, the resulting signal quality is largely invariant to a particular makeup of the interfering signal and depends mainly on the total power and the spectral composition of the interference in the passband of interest. On the other hand, a nonlinear filter is capable of disproportionately affecting spectral densities of signals with distinct temporal and/or amplitude structures. Thus properly implemented nonlinear filtering enables in-band, real-time mitigation of interference with distinct outlier components to levels unattainable by linear filters.

\begin{figure}[!b]
\centering{\includegraphics[width=8.89cm]{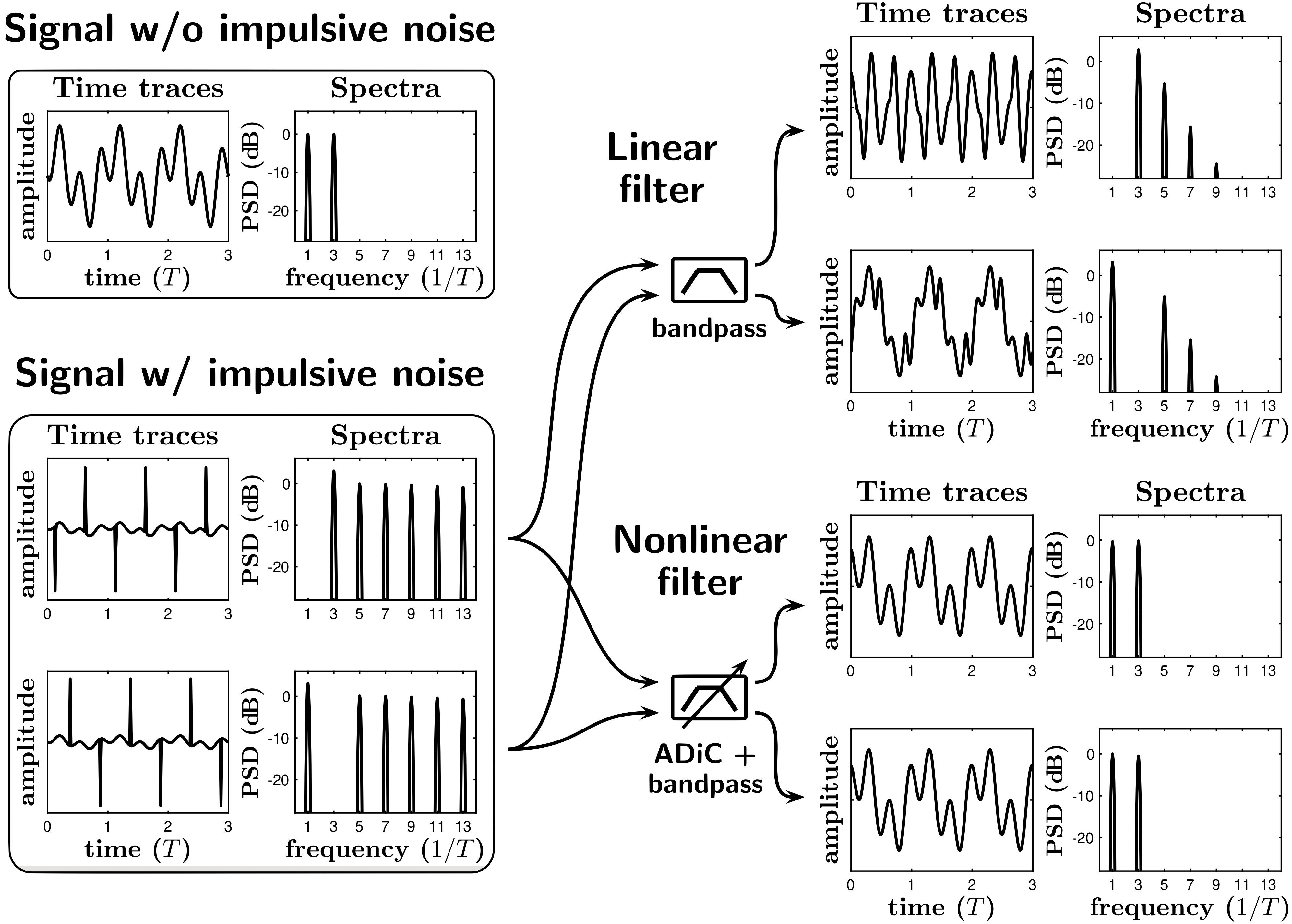}}
\caption{Toy example of suppressing in-band impulsive interference.\label{fig:toy example}}
\end{figure}

This is illustrated by the toy example of Fig.~\ref{fig:toy example}. As shown in the upper left of the figure, the signal of interest consists of a sum of two harmonic tones of equal power, with the periods~$T$ and~$T/3$. In the lower left, a periodic impulse train with the period~$T$ is added to the signal, and the powers of the 1st and the 3rd harmonics of the impulse train are equal to those of the signal. In the first case, the pulse train interferes destructively with the first tone of the signal, and constructively with the second tone. In the second case, the pulse train interferes destructively with the second tone of the signal, and constructively with the first tone.

When a linear filter is used to suppress this impulsive interference while letting through the signal of interest, it neither restores the ``missing" tone nor reduces the power of the signal's tone affected by the constructive interference. This is shown in the upper right of the figure. Here, the bandpass filter consist of a 2nd~order highpass Butterworth filter with the cutoff frequency~$T^{-1}\!/6$, cascaded with a 4th~order lowpass Butterworth filter with the cutoff frequency~$9T^{-1}\!/2$.

On the other hand, as illustrated in the lower right of the figure, a nonlinear filter ahead of the bandpass filter effectively removes the impulse train, along with all its harmonics, restoring the ``missing" tone and providing a two-tone output nearly identical to the signal of interest. Here, a particular nonlinear filter is used (indicated as ``ADiC"), described further in this paper and in~\cite{Nikitin18ADiC-ICC, Nikitin19ADiCpatent}.

\begin{figure}[!b]
\centering{\includegraphics[width=8.89cm]{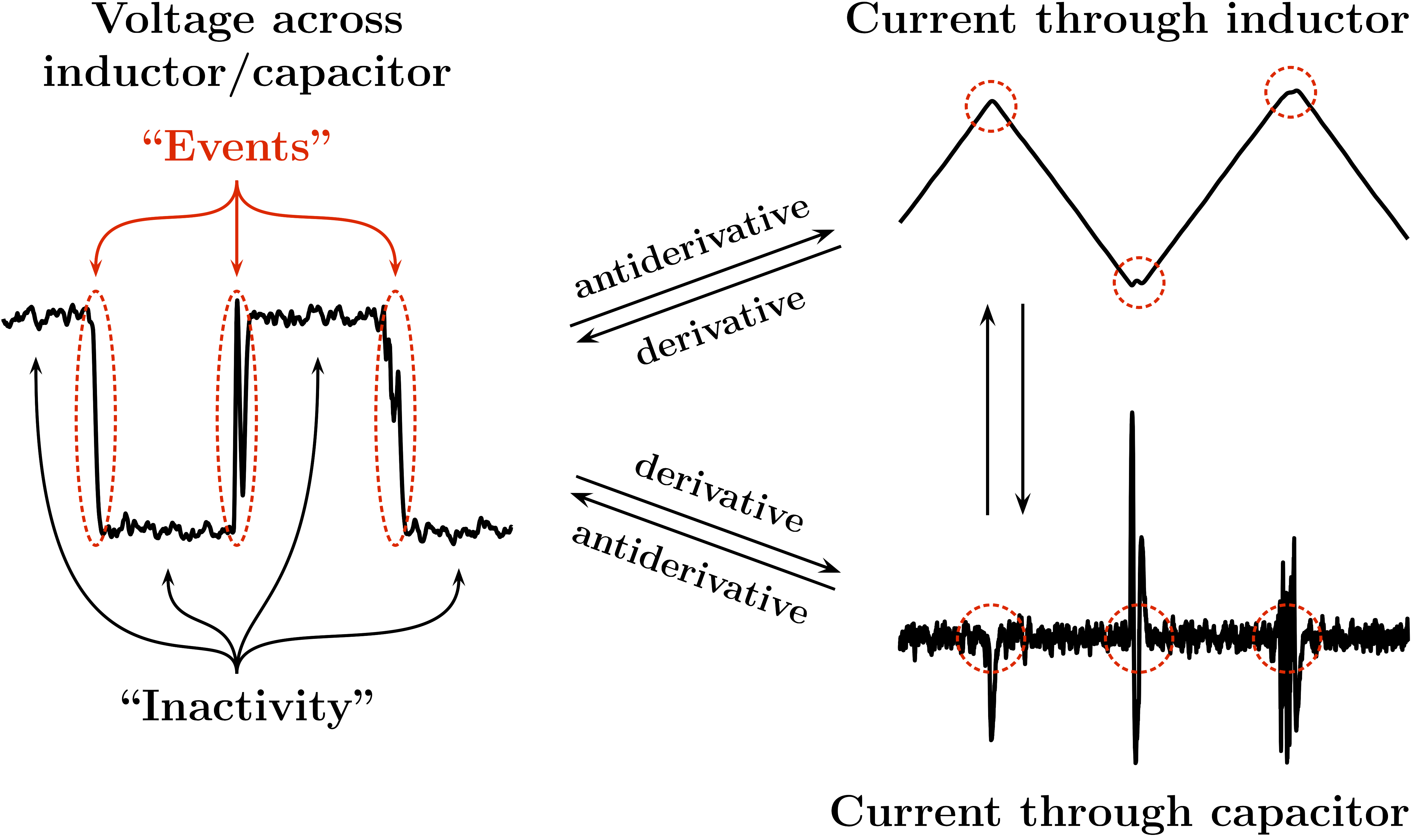}}
\caption{Outlier noise caused by ``events" separated by ``inactivity".\label{fig:origins}}
\end{figure}
\begin{figure*}[!t]
\centering{\includegraphics[width=16.4cm]{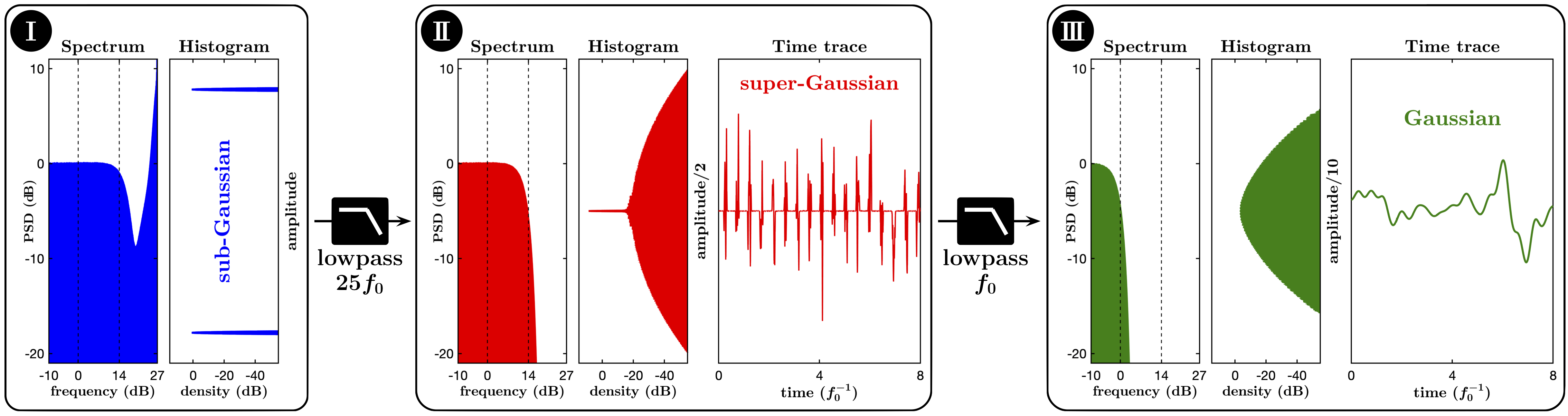}}
\caption{Effect of filtering on amplitude distribution.
\label{fig:modifiable}}
\end{figure*}

\subsection{Omnipresence of outlier noise} \label{subsec:omnipresence}
In addition to ever-present thermal noise, various communication and sensor systems can be affected by interfering signals that originate from a multitude of other natural and technogenic (man-made) phenomena. Such interfering signals often have intrinsic temporal and/or amplitude structures different from the Gaussian structure of the thermal noise. Specifically, interference can be produced by some ``countable" or ``discrete," relatively short duration events that are separated by relatively longer periods of inactivity. Provided that the observation bandwidth is sufficiently large relative to the rate of these non-thermal noise generating events, and depending on the noise coupling mechanisms and the system's filtering properties and propagation conditions, such noise may contain distinct amplitude outliers when observed in the time domain. The presence of different types of such outlier noise is widely acknowledged in multiple applications, under various general and application-specific names, most commonly as {\it impulsive\/}, {\it transient\/}, {\it burst\/}, or {\it crackling\/} noise.

Fig.~\ref{fig:origins} illustrates how a combination of ``events" separated by ``inactivity" can give rise to distinct outliers in the time-domain amplitudes. The transitions between the levels in the square wave are ``events," and the intervals between these transitions are ``inactivity". In particular, in Fig.~\ref{fig:origins} the square wave can be viewed as a voltage across a circuit component. If this component is an inductor, then the current through this component will be a triangle wave (i.e. an antiderivative of the square wave). If the component is a capacitor, then the current through this component will be an impulse train (i.e. a derivative of the square wave). Therefore, although the level transitions in the square wave are not amplitude outliers, they are a source of ``hidden" outlier noise as they can appear as outliers after the square wave is modified, e.g. by linear filtering. In the analog domain, such filtering can be viewed as a linear combination of the signal with its derivatives and antiderivatives (e.g. convolution) of various orders. In the digital domain, it is a combination of differencing and summation operations.

Examples of outlier noise arising from natural phenomena include ice cracking (in polar regions) and snapping shrimp (in warmer waters) affecting underwater acoustic communications and sonar, or lightning discharges in thunderstorms affecting RF systems. However, the most prevalent source of outlier noise generating events is that of technogenic origin. In particular, the following is a simplified explanation of the outlier nature of technogenic noise produced by digital electronics and communication systems. An idealized discrete-level (digital) signal can be viewed as a linear combination of Heaviside unit step functions~\cite{Bracewell2000Fourier}. Since the derivative of the Heaviside unit step function is the Dirac $\delta$-function~\cite{Dirac58principles}, the derivative of an idealized digital signal is a linear combination of Dirac $\delta$-functions, which is a limitlessly impulsive signal with zero interquartile range and infinite peakedness. The derivative of a ``real" (i.e. no longer idealized) digital signal can thus be viewed as a convolution of a linear combination of Dirac $\delta$-functions with a continuous kernel. If the kernel is sufficiently narrow (for example, the bandwidth is sufficiently large), the resulting signal will appear as a transient pulse train protruding from a disperse background. Hence outlier electromagnetic interference (EMI) is inherent in digital electronics and communication systems, transmitted into a system in various ways, including electrostatic coupling, electromagnetic induction, or RF radiation. For example, descriptions of detailed mechanisms of impulsive nature of out-of-band and adjacent-channel interference in digital communication systems can be found in~\cite{Nikitin11aRWS, Nikitin11bEURASIP, Nikitin15OOB}.

\subsection{Outlier noise detection and mitigation} \label{subsec:detection and mitigation}
Although the detrimental effects of EMI are broadly acknowledged in the industry, its outlier nature often remains indistinct, and its omnipresence and impact, and thus the potential for its mitigation, remain greatly underappreciated. This paper provides an overview of the methodology and tools for real-time mitigation of outlier noise in general and ``hidden" wideband outlier noise in particular. Such mitigation is performed as a ``first line of defense" against interference ahead of, or in the process of, reducing the bandwidth to that of the signal of interest. Either used by itself, or in combination with subsequent interference mitigation techniques, this approach provides interference mitigation levels otherwise unattainable, with the effects, depending on particular interference scenarios, ranging from ``no harm" to substantial.

In the next section we explain why underlying outlier nature of interference often remains obscured, and demonstrate how its out-of-band observation enables its in-band mitigation. We then introduce the basic real-time nonlinear components that are used in outlier noise filtering, and give examples of their implementation. We further describe complete intermittently nonlinear filtering arrangements for wide- and narrow-band outlier noise reduction, and provide several illustrations of their performance and the effect on channel capacity. Penultimately, we outline and illustrate ``effectively analog" digital implementation of these filtering structures, including their modifications for addressing complex practical interference scenarios. Finally, we briefly discuss broader applications of these nonlinear filtering techniques, discuss the ongoing development of the platform for their demonstration and testing, and outline the direction of future work.

\section{Elusive nature of outlier noise} \label{sec:elusive nature}
There are two fundamental reasons why the outlier nature of many technogenic interference sources is often dismissed as irrelevant. The first one is a simple lack of motivation.  As discussed in the introduction, without using nonlinear filtering techniques the resulting signal quality is largely invariant to a particular time-amplitude makeup of the interfering signal and depends mainly on the total power and the spectral composition of the interference in the passband of interest. Thus, unless the interference results in obvious, clearly identifiable outliers in the signal's band, the ``hidden" outlier noise does not attract attention.

The second reason is highly elusive nature of outlier noise, and inadequacy of tools used for its consistent observation and/or quantification. For example, neither power spectral densities (PSDs) nor their short-time versions (e.g. spectrograms) allow us to reliably identify outliers, as signals with very distinct temporal and/or amplitude structures can have identical spectra. Amplitude distributions (e.g. histograms) are also highly ambiguous as an outlier-detection tool. Although a super-Gaussian (heavy-tailed) amplitude distribution of a signal does normally indicate presence of outliers, it does not necessarily reveal presence or absence of outlier noise in a wideband signal. Indeed, a wide range of powers across a wideband spectrum allows a signal containing outlier noise to have any type of amplitude distribution. More important, the amplitude distribution of a non-Gaussian signal is generally modifiable by linear filtering (e.g. see Fig.~\ref{fig:origins}), and such filtering can often convert the signal from sub-Gaussian into super-Gaussian, and {\it vice versa\/}~\cite{Nikitin13adaptive, Nikitin15OOB}. Therefore apparent outliers in a signal can disappear and reappear due to various filtering effects, including fading and multipass, as the signal propagates through media and/or the signal processing chain.

This is illustrated in Fig.~\ref{fig:modifiable}, where the original wideband signal is the ``raw" output of a 1-bit $\Delta\Sigma$ modulator given a ``bursty" input (Panel~I). It is clearly a two-level signal with a sub-Gaussian amplitude distribution, and may also represent a bi-stable process in general. Specifically, the sampling rate in this example is~$10^3f_0$. Panel~II shows the output of a 1st~lowpass filter, with the cutoff~$25f_0$, applied to the 1-bit signal. This output is a bursty train with a super-Gaussian amplitude distribution and distinct ``bursty" outliers. In Panel~III, the output of the 1st~lowpass filter is further filtered with a 2nd~lowpass filter with the cutoff~$f_0$. Now there are no apparent outliers in the output, and its amplitude distribution is effectively Gaussian. Thus even simple reduction in bandwidth of a signal by lowpass filtering can make apparent outliers wax and wane.
Further examples can be found in~\cite{Nikitin15OOB} and~\cite{Nikitin13adaptive}.

\subsection{What hides outlier noise?} \label{subsec:what hides}
\begin{figure}[!t]
\centering{\includegraphics[width=8.6cm]{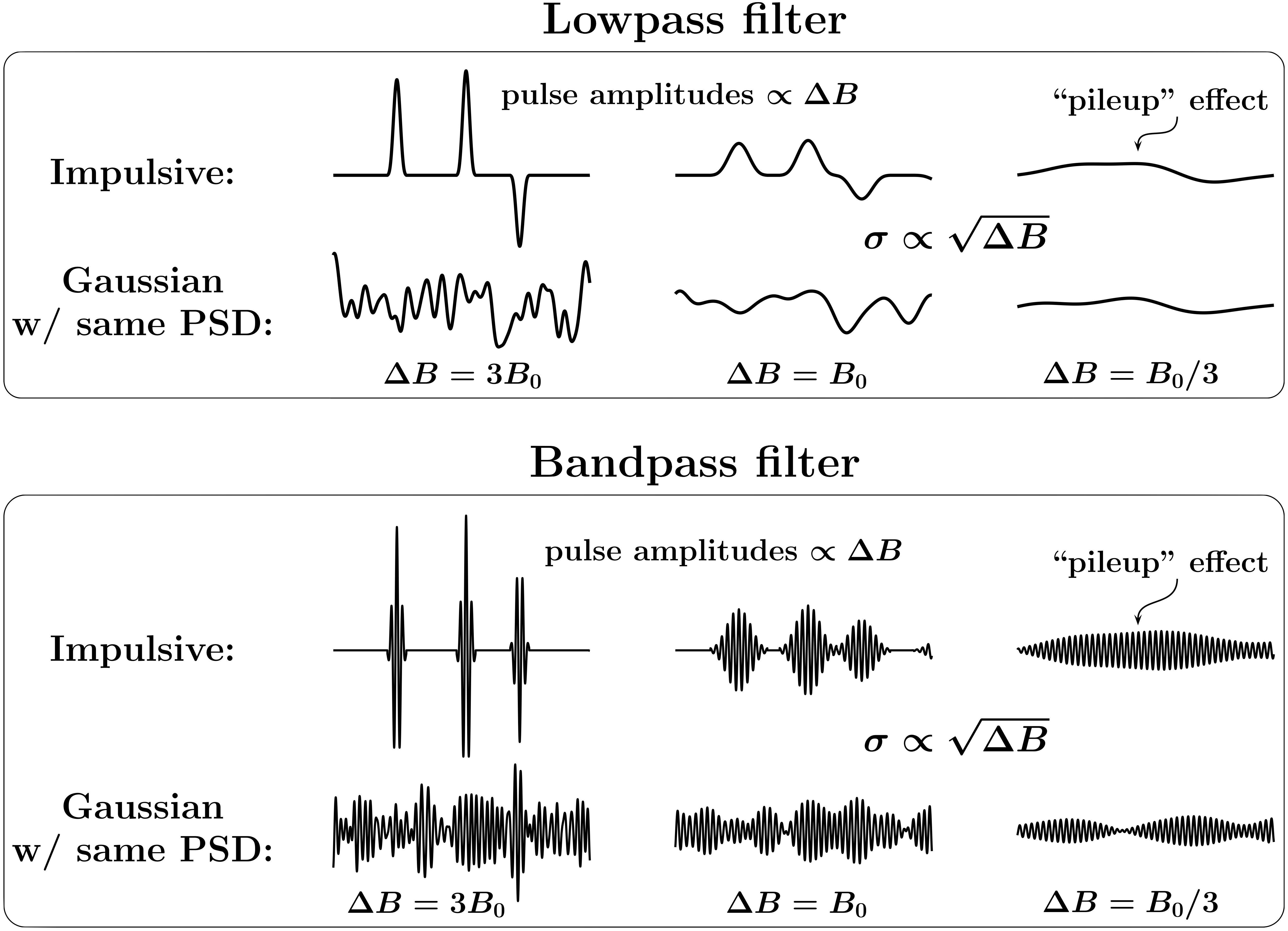}}
\caption{Reduction in bandwidth ``hides" outlier noise.\label{fig:elusive}}
\end{figure}

The example given in Fig.~\ref{fig:modifiable} also demonstrates that, {\it once outlier noise becomes apparent\/}, additional reduction in bandwidth typically makes it less evident. Fig.~\ref{fig:elusive} further illustrates the basic mechanism of outlier noise ``disappearance" with the reduction in observation bandwidth.

First note that a band-limited signal will not be affected by the change in the bandwidth of a filter, as long as the filter does not attenuate the signal's frequencies. Hence Fig.~\ref{fig:elusive} compares the effects of reducing the bandwidth of a lowpass or a bandpass filter only on Gaussian and impulsive noise. When the bandwidth~$\Delta{B}$ is reduced, the standard deviation of the noise decreases as a square root of its bandwidth, ${\sigma\propto\sqrt{\Delta{B}}}$. For Gaussian noise, its standard deviation is proportional to its amplitude. The amplitude of the impulsive noise, however, is affected differently by the bandwidth change. For example, as shown in Fig.~\ref{fig:elusive}, the amplitudes of the stand-alone pulses that effectively represent the impulse responses of the respective filters decrease proportionally to their bandwidth, faster than the amplitude of the Gaussian noise. Thus the bandwidth reduction causes the impulsive noise to protrude less from the Gaussian background. On the other hand, the width of the pulses is inversely proportional to the bandwidth. When the width of the pulses becomes greater than the distance between them, the pulses begin to overlap and interfere with each other. For a random pulse train, when the ratio of the bandwidth and the pulse arrival rate becomes significantly smaller than the time-bandwidth product of a filter, the resulting signal becomes effectively Gaussian due to the so-called ``pileup effect"~\cite[e.g.]{Nikitin98ppileup}, making the impulsive noise completely disappear.

\begin{figure}[!t]
\centering{\includegraphics[width=8.6cm]{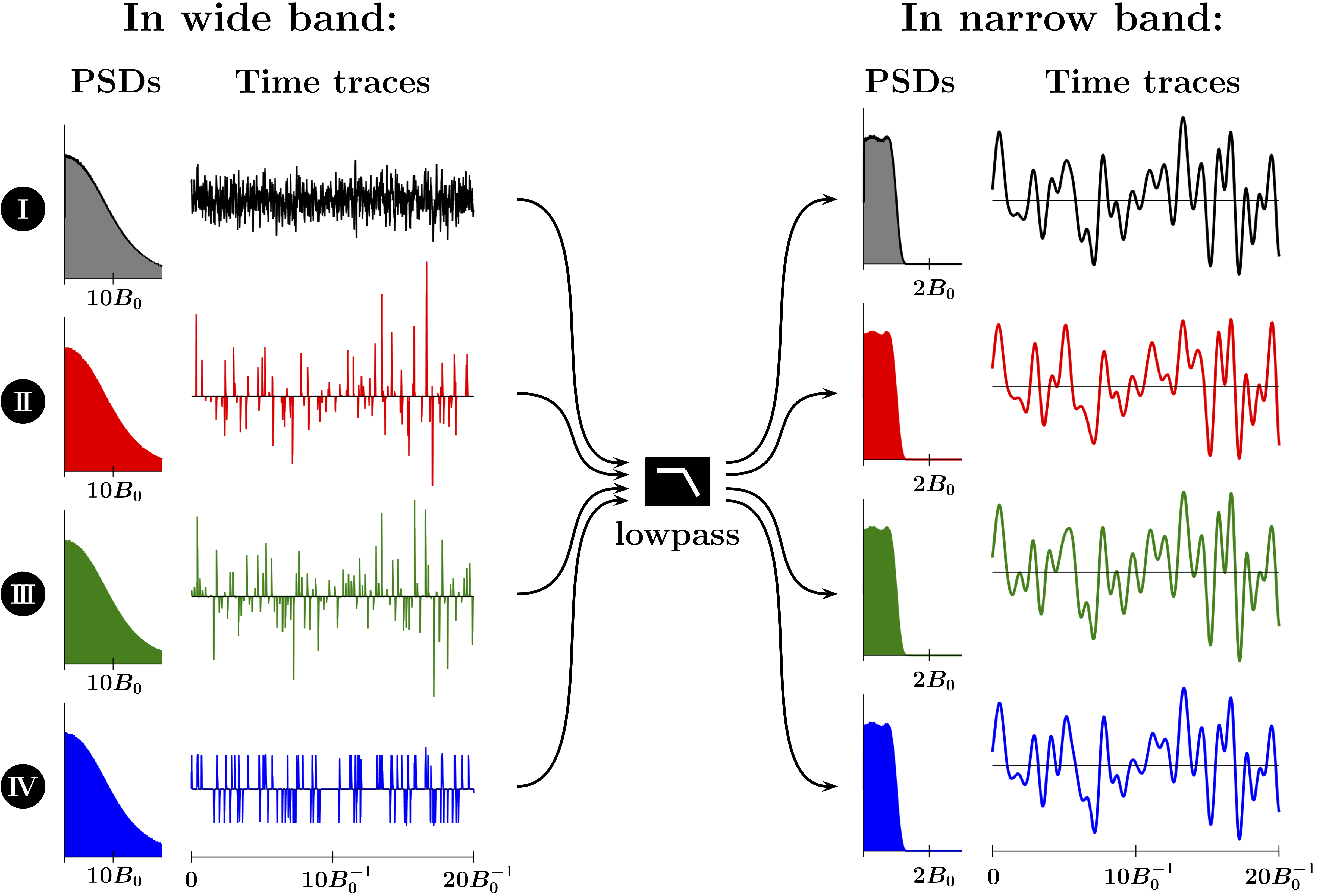}}
\caption{Hidden vs. apparent outlier noise.\label{fig:hidden}}
\end{figure}

Fig.~\ref{fig:hidden} provides an additional illustration of importance of excess bandwidth for outlier noise detection and mitigation. In the figure, all shown noises have identical PSDs and are intentionally constructed to have very similar, and effectively Gaussian, time-domain appearances in the narrow band. However, they all have very different time-amplitude structures, which is clearly visible in the wideband time traces. As discussed earlier (and unlike the wideband Gaussian noise shown in panel~I), the wideband noises with outlier structures shown in panels~II, III,~and~IV are {\it mitigable by nonlinear filtering\/}.

\subsection{``Outliers" vs. ``outlier noise"} \label{subsec:outliers}
Even when sufficient excess bandwidth is available for outlier noise observation, outlier noise mitigation faces significant challenges when the typical amplitude of the noise outliers is not significantly larger than that of the signal of interest. That would be the case, e.g., if the signal of interest itself contains strong outliers, or for large signal-to-noise ratios (SNRs), especially when combined with high rates of the noise-generating events. In those scenarios removing outliers from the signal+noise mixture may degrade the signal quality instead of improving it. This is illustrated in Fig.~\ref{fig:outliers}.

The left-hand side of the figure shows a fragment of a low-frequency signal affected by a wideband noise containing outliers. However, the amplitudes of the signal and the noise outliers are such that only one of the outlier noise pulses is an outlier for the signal+noise mixture. The right-hand side of the figure illustrates that removing only this outlier increases the baseband noise, instead of decreasing it by the ``outlier noise" removal.

\subsection{``Excess band" observation for in-band mitigation} \label{subsec:excess band}
\begin{figure}[!b]
\centering{\includegraphics[width=8.6cm]{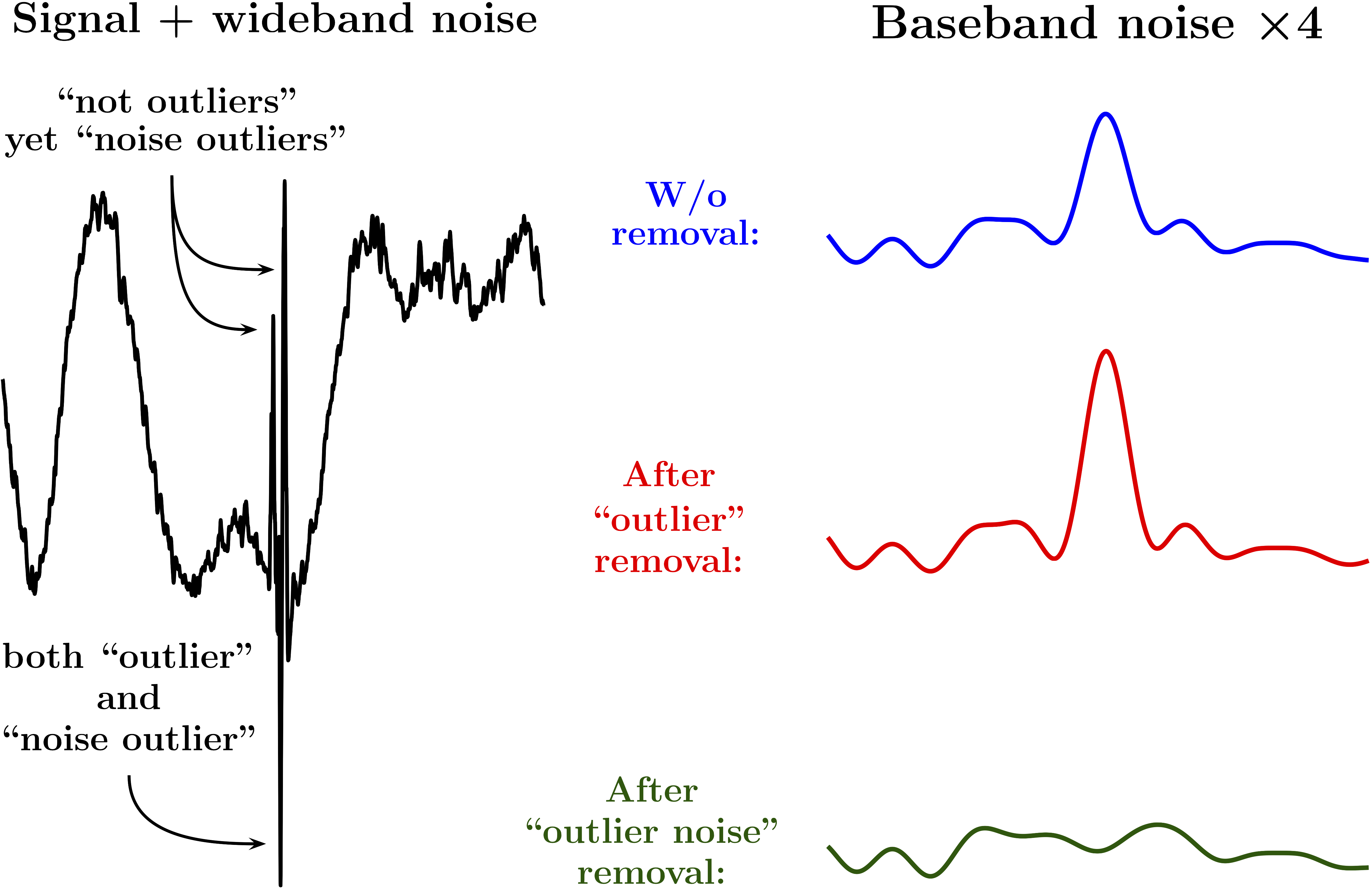}}
\caption{Removing ``outliers" instead of ``outlier noise" degrades signal.\label{fig:outliers}}
\end{figure}

\begin{figure}[!t]
\centering{\includegraphics[width=8.6cm]{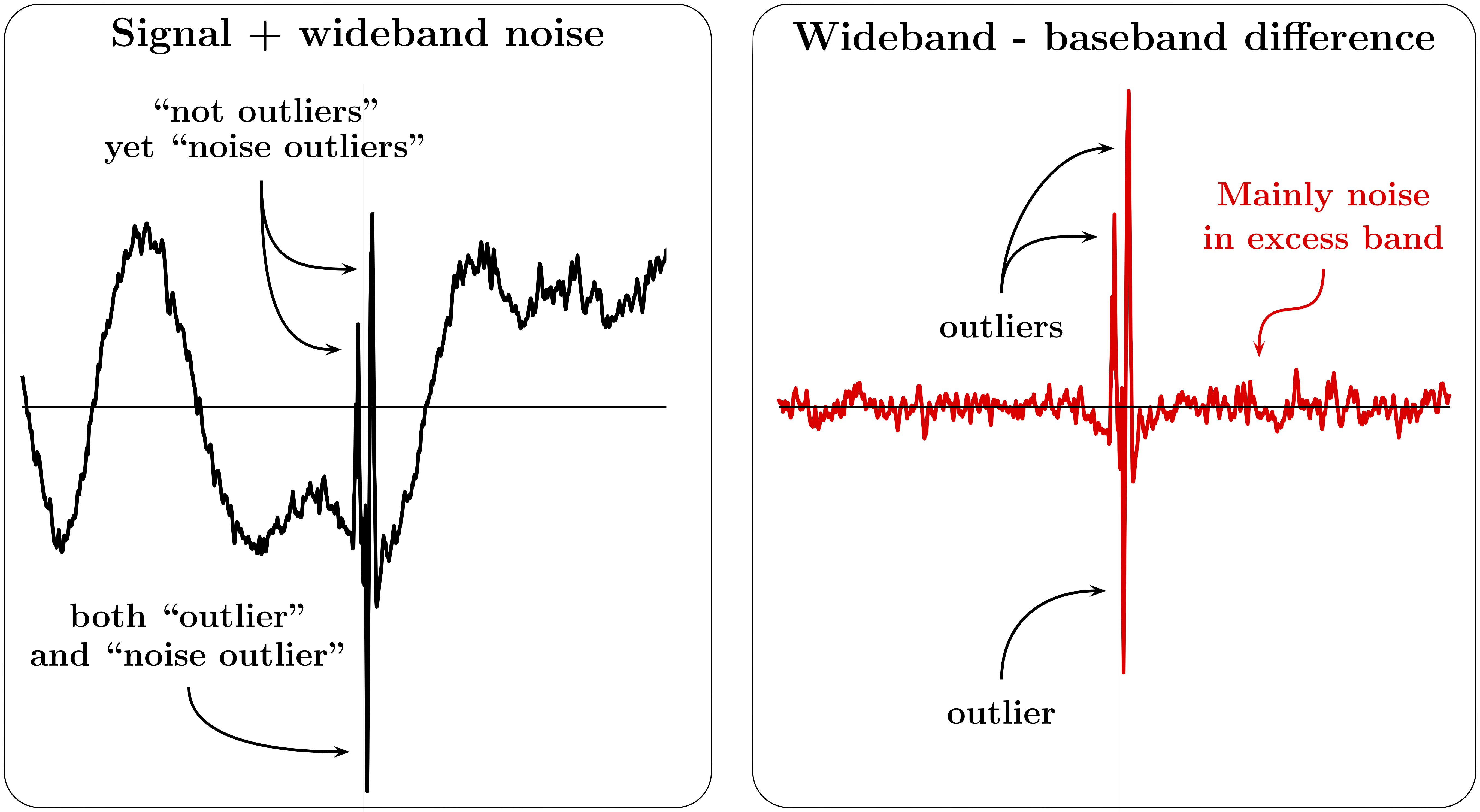}}
\caption{``Excess band" observation of outlier noise.\label{fig:difference signal}}
\end{figure}

As discussed in~Section~\ref{subsec:what hides}, a linear filter affects the amplitudes of the signal of interest, wideband Gaussian noise, and wideband outlier noise differently. Fig.~\ref{fig:difference signal} illustrates how we can capitalize on these differences to reliably distinguish between ``outliers"   and ``outlier noise". The left-hand side of the figure shows the same fragment of the low-frequency signal affected by the wideband noise containing outliers as in the example of Fig.~\ref{fig:outliers}. This signal+noise mixture can be viewed as an output of a wideband front-end filter. When applied to the output of the front-end filter, a baseband lowpass filter that does not attenuate the low-frequency signal will still significantly reduce the amplitude of the wideband noise. Then the difference between the input signal+noise mixture and the output of the baseband filter with zero group delay across signal's band will mainly contain the wideband noise filtered with highpass filter obtained by spectral inversion of the baseband filter. This is illustrated in the right-hand side of Fig.~\ref{fig:difference signal}, showing that now the outliers in the difference signal are also the noise outliers.


\begin{figure}[!t]
\centering{\includegraphics[width=8.6cm]{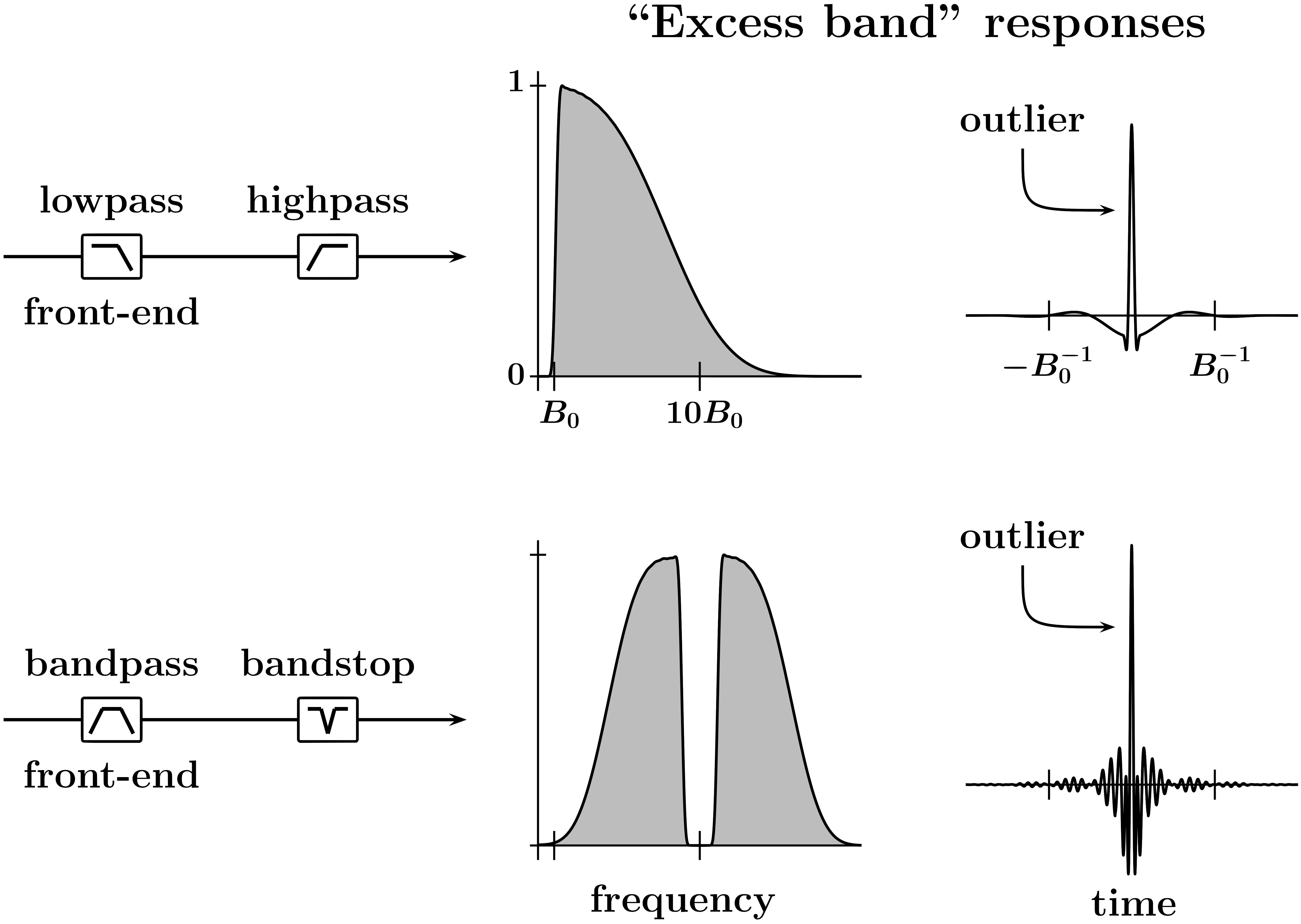}}
\caption{Illustrative examples of excess band responses.\label{fig:excess band}}
\end{figure}

Therefore, detection of outlier noise can be accomplished by an ``excess band filter" constructed as a cascaded lowpass/highpass (for a baseband signal of interest), or as a cascaded bandpass/bandstop filter (for a passband signal of interest). This is illustrated in Fig.~\ref{fig:excess band} where, for simplicity, finite impulse response (FIR) filters are used. Provided that the ``excess band" is sufficiently wide in comparison with the band of the signal of interest, the impulse response of an excess band filter contains a distinct outlier component. When convolved with a band-limited signal affected by a wideband outlier noise, such a filter will suppress the signal of interest while mainly preserving the outlier structure of the noise. In Section~\ref{sec:tools} we show how such excess band observation of outlier noise enables its efficient in-band mitigation.

\subsection{``Peakedness" for assessing mitigation potential} \label{subsec:peakedness}
\begin{figure}[!b]
\centering{\includegraphics[width=8.6cm]{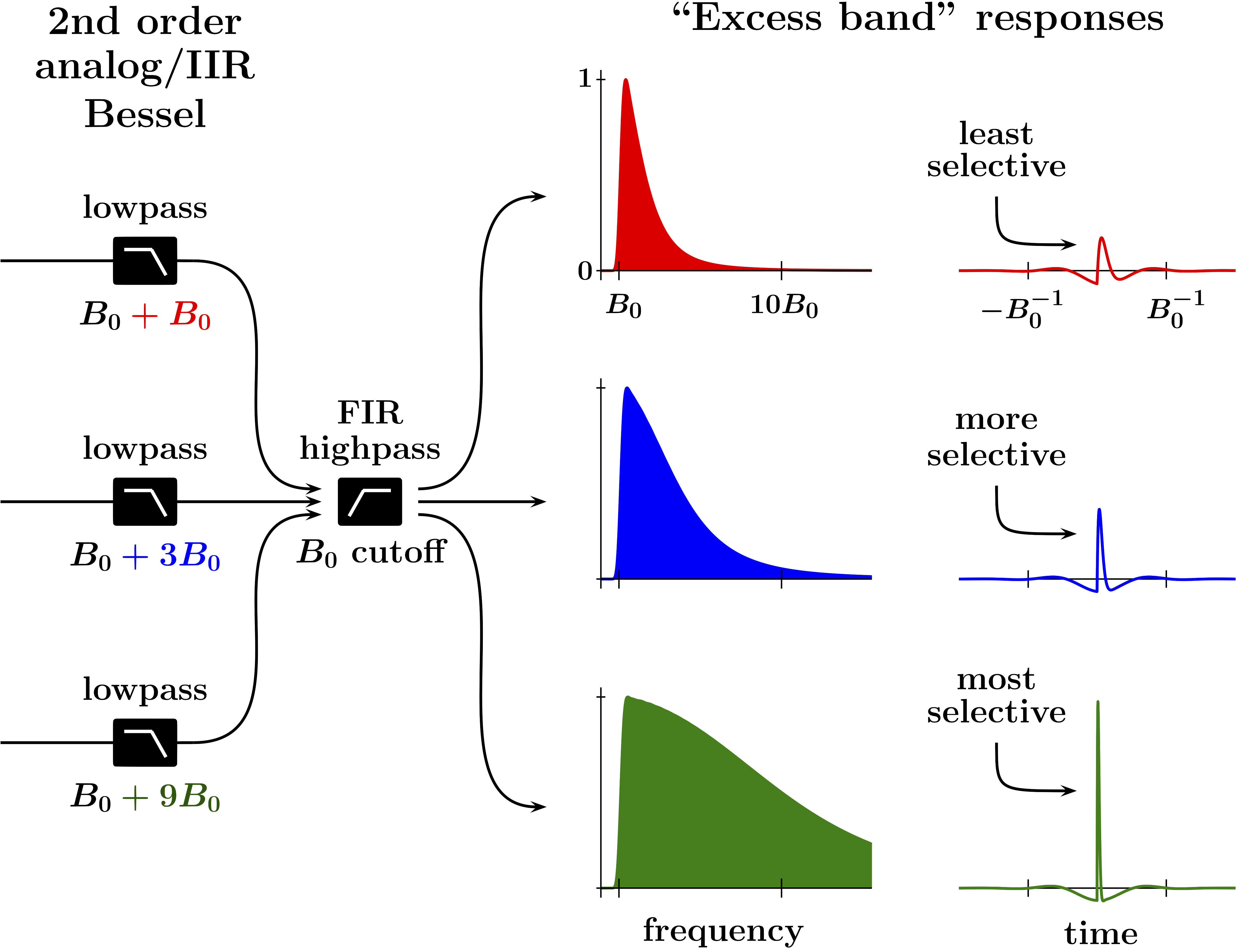}}
\caption{Effect of bandwidth on excess band selectivity.\label{fig:excess bandwidth}}
\end{figure}

The amount of excess bandwidth that can be allocated for outlier noise mitigation depends on the particular requirements and constraints placed on a system, and the excess bandwidth availability affects both the ``mitigable rates" (e.g. in terms of the rates of outlier generating events) and ``mitigable SNRs" (e.g. in terms of outlier-to-thermal noise powers) of the outlier noise. Fig.~\ref{fig:excess bandwidth} provides a qualitative illustration of how the increase in the bandwidth of the front-end filter affects selectivity of the excess band.

\begin{figure}[!t]
\centering{\includegraphics[width=8.6cm]{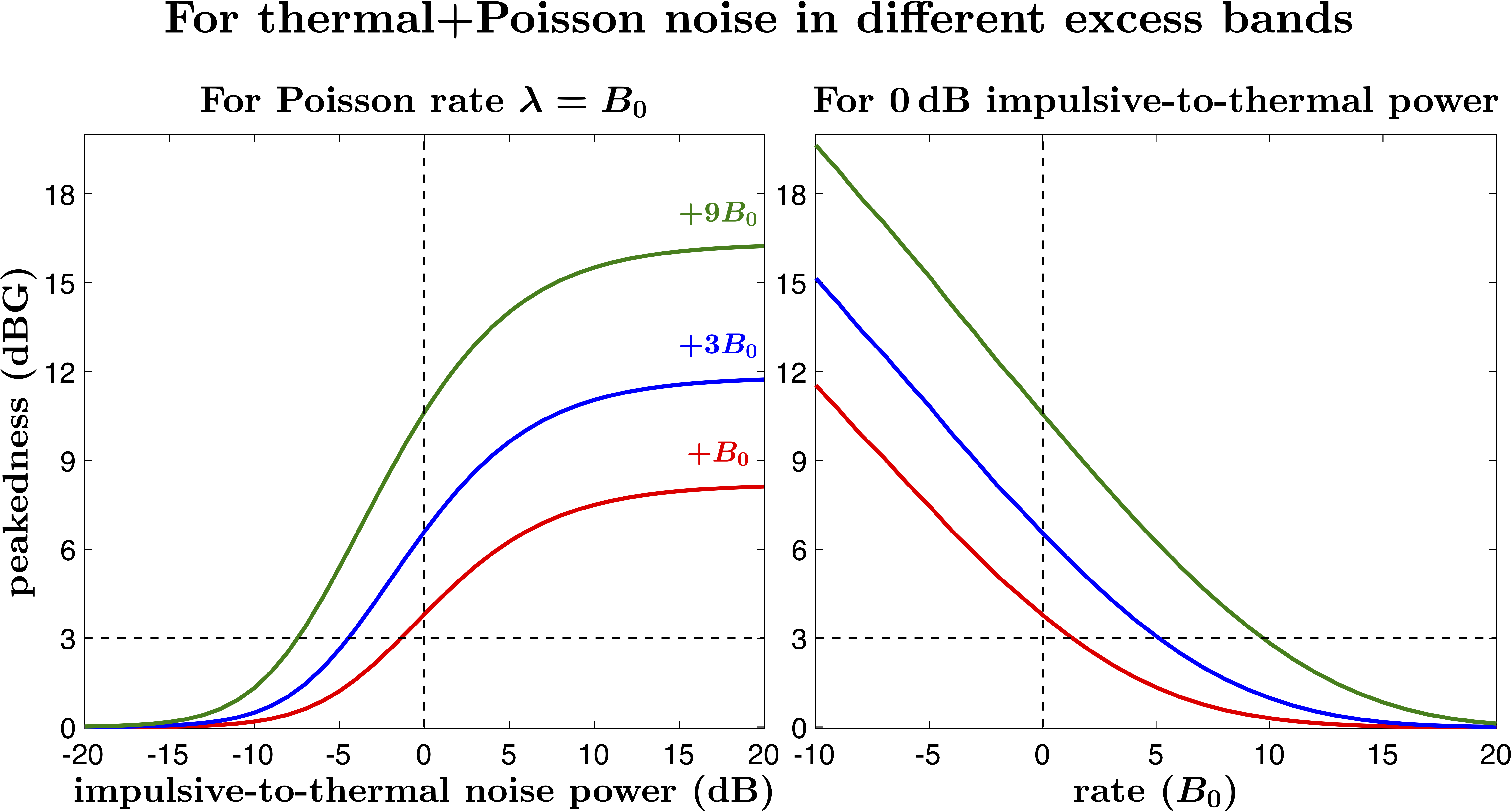}}
\caption{Peakedness for assessing mitigation potential.\label{fig:peakedness}}
\end{figure}

A useful quantifier of the prevalence of noise outliers, and thus of the potential for their mitigation, can be the peakedness of the noise relative to a Gaussian (aka normal) distribution. Based on the definition of kurtosis in~\cite{MILCOM13:Abramowitz72handbook}, the peakedness of a real signal~$x(t)$ can be measured in units of ``decibels relative to Gaussian" (dBG) as follows~\cite{Nikitin13adaptive}:
\beginlabel{equation}{eq:kurtosis dBG}
  K_{\rm dBG}(x) = 10\lg \textstyle
  \left[ \frac{\left\langle \left(x - \langle x\rangle\right)^4
  \right\rangle} {3\left\langle \left(x - \langle x\rangle\right)^2 \right\rangle^2 } \right],
\end{equation}
where the angular brackets denote time averaging. According to this definition, a normal distribution has zero dBG peakedness, while sub-Gaussian and super-Gaussian distributions have negative and positive dBG peakedness, respectively. In terms of the amplitude distribution of a signal, a higher peakedness compared to a Gaussian distribution (super-Gaussian) normally translates into ``heavier tails" than those of a Gaussian distribution. In the time domain, high peakedness implies more frequent occurrence of outliers. Note that, while positive dBG peakedness indicates the presence of an outlier component in a signal, negative or zero dBG peakedness does not necessarily exclude the presence of outliers. As follows from the linearity property of kurtosis, a mixture of super-Gaussian (positive kurtosis) and sub-Gaussian (negative kurtosis) signals can have any intermediate value of kurtosis. For example, the peakedness of an ideal square, triangle, and sine wave is approximately $-4.77$, $-2.22$, and $-3.01$\,dBG, respectively, and these waveforms can absorb significant amounts of outlier noise before their peakedness becomes positive. However, for a mixture of a thermal and an outlier noise positive peakedness does indicate presence of outliers, and it can be used for assessing the outlier noise mitigation potential. Fig.~\ref{fig:peakedness} provides examples of peakedness, as functions of the impulsive-to-thermal noise power and of the pulse rate, of a wideband Poisson noise with normally distributed amplitudes filtered with the excess band responses shown in Fig.~\ref{fig:excess bandwidth}. This demonstrates that higher excess bandwidth leads to both higher mitigable rates and higher mitigable SNRs of outlier noise.


\section{Methodology and tools for outlier noise mitigation} \label{sec:tools}
Once both the concept of ``hidden" outlier interference and the nonlinear tools for its mitigation are entered into consideration, the signal processing part of the overall approach to interference reduction can be briefly outlined as follows:
First, we remove the wideband outlier noise, while preserving the signal of interest and the wideband non-outlier noise that is not removable by nonlinear filtering, e.g. the thermal noise. Such outlier noise removal should be performed either in the analog domain ahead of analog-to-digital conversion (A/D), or in the process of A/D, and it can be done with or without reducing the bandwidth of the input signal to that of the signal of interest~\cite{Nikitin19ADiCpatent, Nikitin18ADiC-ICC}. Next, linear filtering (e.g. matched filtering) can be performed in the digital domain to maximize the SNR. Finally, and based on the {\it a priori\/} knowledge of the signal of interest's structure, the in-band signal outliers can be detected and removed or separated from the rest of the narrow-band signal.

\begin{figure}[!b]
\centering{\includegraphics[width=8.6cm]{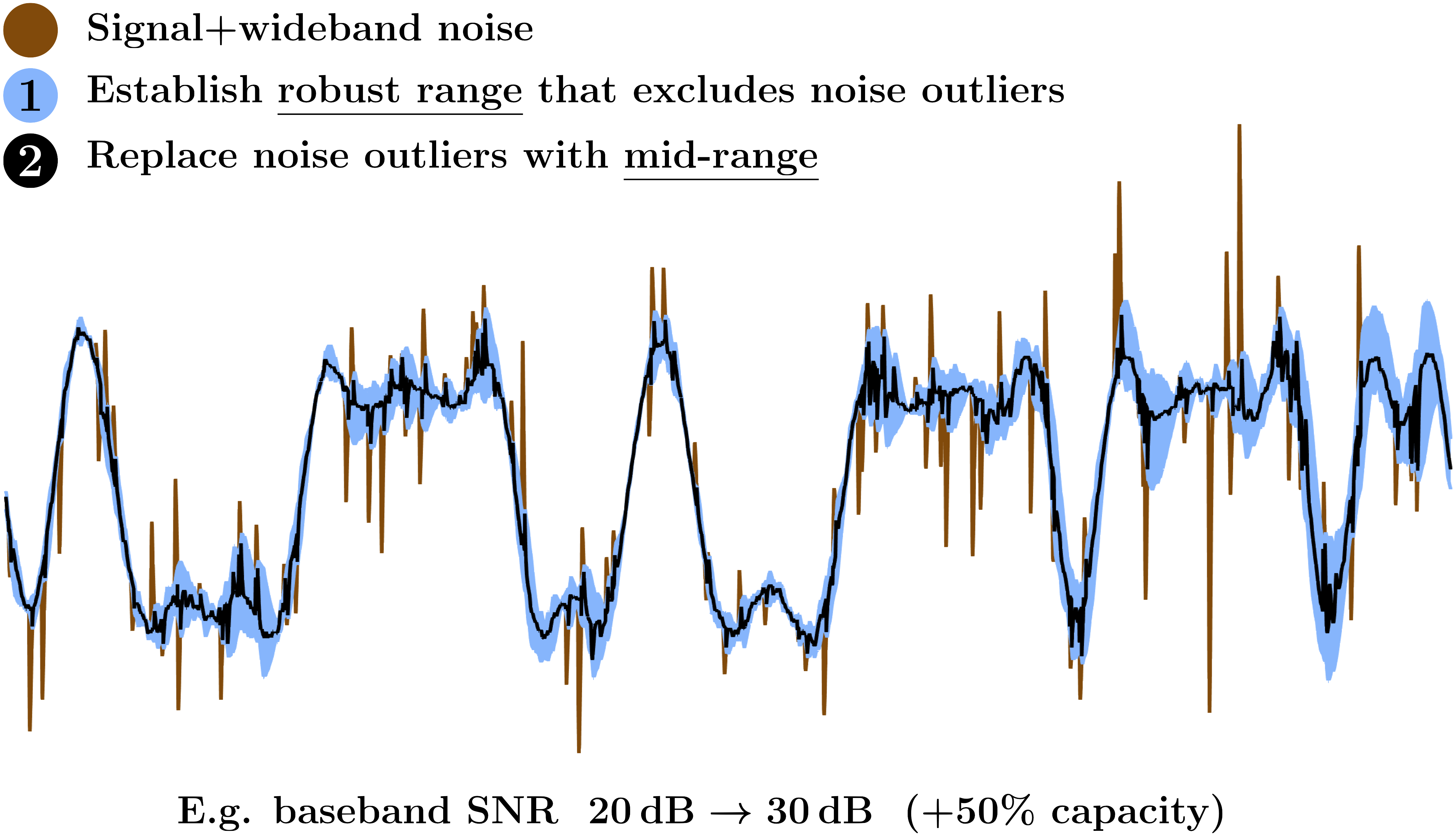}}
\caption{Removing wideband outlier noise while preserving signal of interest.\label{fig:ADiC concept}}
\end{figure}

It is important to emphasize the difference between the wide- and narrow-band outlier noise reduction. Since narrowband outliers are confined to the same frequency band as the signal of interest, a narrowband signal+interference mixture should be treated in a similar fashion to a wideband outlier noise without the signal of interest. On the other hand, as discussed in Section~\ref{sec:elusive nature}, efficient mitigation of wideband outlier noise requires its observation in the ``excess band" so that the signal itself can be mainly excluded. With this in mind, Fig.~\ref{fig:ADiC concept} illustrates the basic concept of wideband outlier noise removal while preserving the signal of interest and the wideband non-outlier noise. First, we establish a robust range that excludes noise outliers while including the signal of interest. Then, we replace the outlier values with those in mid-range. Note that this simplified illustration does not include any strong non-outlier wideband components, e.g. adjacent channel interference. Addressing such more complex interference compositions is briefly discussed in~Section~\ref{subsubsec:complex}.

We would like to mention in passing that, when we are not constrained by the needs for either analog or wideband, high-rate real-time digital processing, in the digital domain the requirements outlined in Fig.~\ref{fig:ADiC concept} can perhaps be satisfied by a {\it Hampel filter\/}~\cite{Hampel74influence} or by one of its variants~\cite{Pearson16generalized}. In a Hampel filter the ``mid-range" is calculated as a windowed median of the input, and the range is determined as a scaled absolute deviation about this windowed median. However, Hampel filtering cannot be performed in the analog domain, and/or it becomes prohibitively expensive in high-rate real-time digital processing.

\begin{figure}[!t]
\centering{\includegraphics[width=8.6cm]{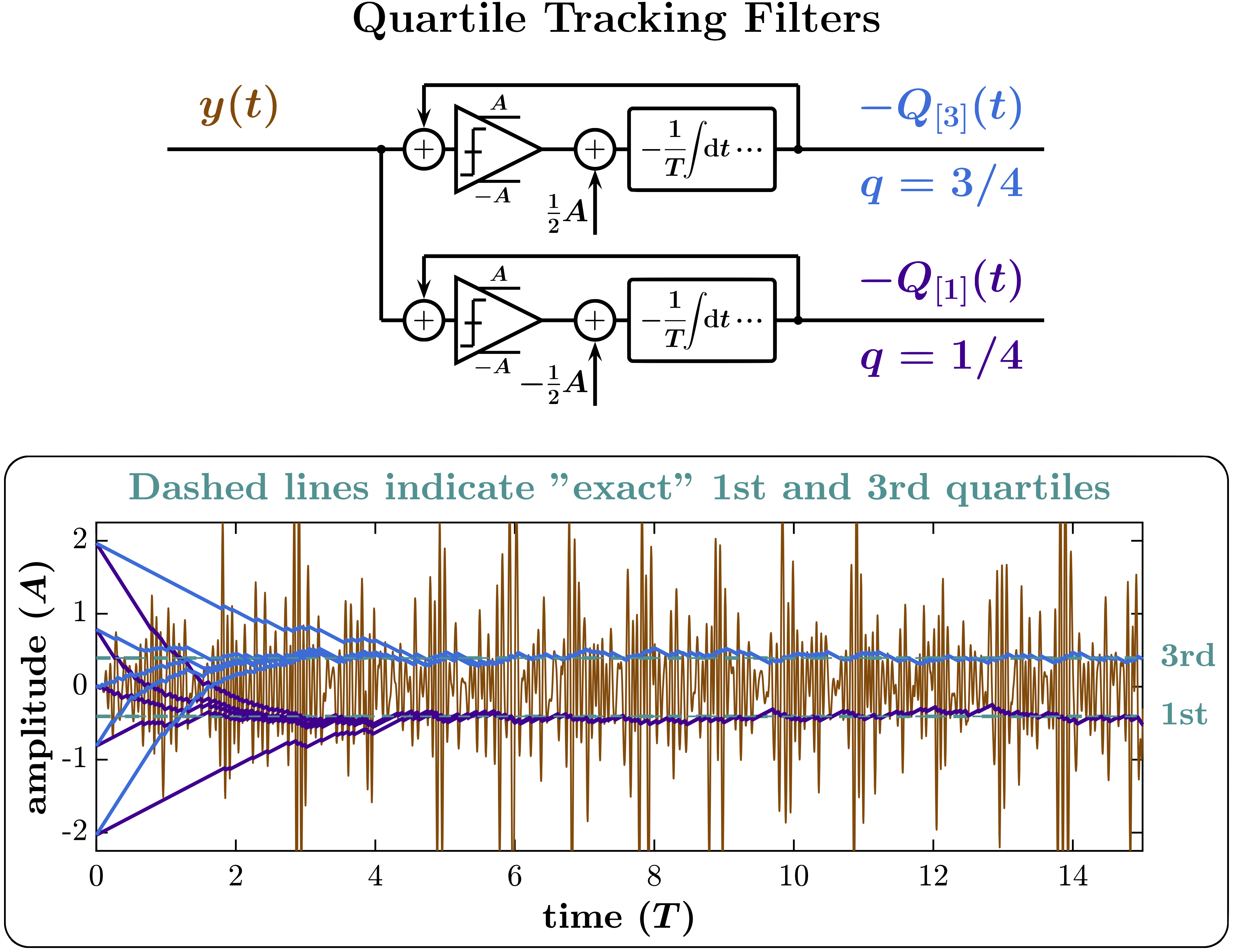}}
\caption{QTFs' convergence to steady state for different initial conditions.
\label{fig:quartile convergence}}
\end{figure}
\begin{figure}[!b]
\centering{\includegraphics[width=8.6cm]{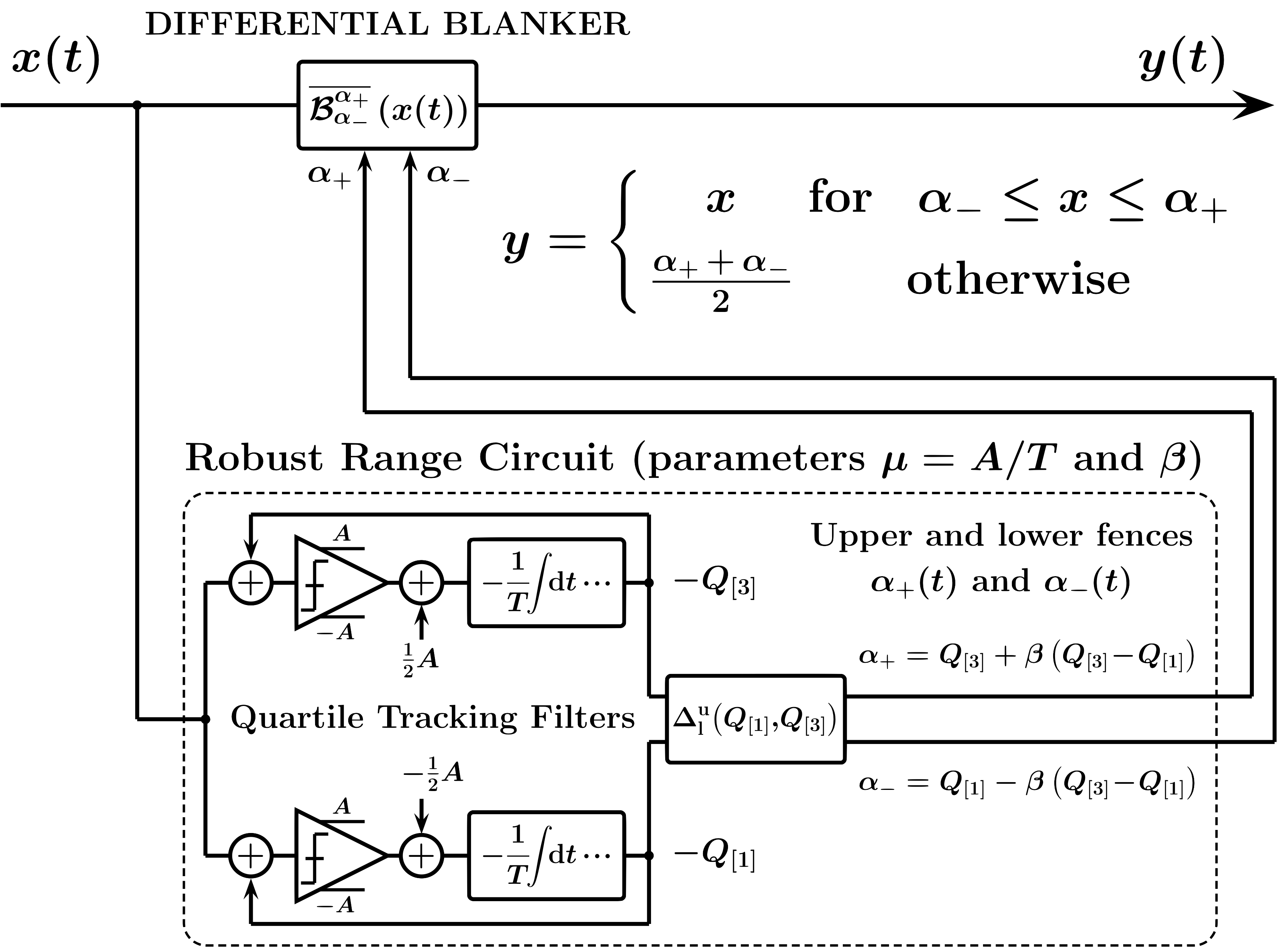}}
\caption{Diagram of basic ADiC structure.
\label{fig:basic ADiC}}
\end{figure}

\subsection{Quantile Tracking Filters for robust range and mid-range} \label{subsec:QTFs}
A robust range ${[\alpha_-,\alpha_+]}$ that excludes outliers of a signal can also be obtained as a range between {\it Tukey's fences\/}~\cite{Tukey77exploratory} constructed as linear combinations of the 1st ($Q_{[1]}$) and the 3rd ($Q_{[3]}$) quartiles of the signal in a moving time window:
{\small
\beginlabel{equation}{eq:Tukey's range}
  [\alpha_-,\alpha_+] = {\big [}Q_{[1]}\!-\!\beta\left(Q_{[3]}\!-\!Q_{[1]}\right)\!,\,Q_{[3]}\!+\!\beta\left(Q_{[3]}\!-\!Q_{[1]}\right)\!{\big ]},
\end{equation}
}
where $\alpha_+$, $\alpha_-$, $Q_{[1]}$, and $Q_{[3]}$ are time-varying quantities, and $\beta$ is a scaling parameter of order unity (e.g. $\beta=1.5$). In practical analog and/or real-time digital implementations, approximations for the time-varying quartile values can be obtained by means of Quantile Tracking Filters (QTFs) introduced in~\cite{Nikitin17nonlinear, Nikitin18ADiC-ICC} and described in detail in~\cite{Nikitin19ADiCpatentCIPs}. In brief, the signal~$Q_q(t)$ that is related to a given input~$y(t)$ by the equation
\beginlabel{equation}{eq:Qq}
  \frac{\d}{\d{t}}\, Q_q = \frac{A}{T}\, \left[\sgn(y\!-\!Q_q) + 2q-1\right],
\end{equation}
where~$A$ is a parameter with the same units as~$y$ and $Q_q$, and~$T$ is a constant with the units of time, can be used to approximate (``track") the $q$-th~quantile of $y(t)$ for the purpose of establishing a robust range ${[\alpha_-,\alpha_+]}$. (See~\cite{Nikitin03signal, Nikitin04adaptive} for discussion of quantiles of continuous signals.) For example, Fig.\,\ref{fig:quartile convergence} illustrates, for a particular input~$y(t)$, the QTFs' convergence to steady states for different initial conditions.

Linear combinations of QTF outputs can also be used to establish the mid-range that replaces the outlier values. For example, the signal values that protrude from the range ${[\alpha_-,\alpha_+]}$ can be replaced by the output of a {\it Trimean Tracking Filter\/} (TTF) $(Q_{[1]}+wQ_{[2]}+Q_{[3]})/(w+2)$, where~$w\ge 0$ (e.g. {\it Tukey's trimean\/} for $w=2$)~\cite{Tukey77exploratory, Nikitin19ADiCpatentCIPs}. Then such mid-range level can be called a {\it Differential Clipping Level\/} (DCL), and a filter that established the range~${[\alpha_-,\alpha_+]}$ and replaces outliers with the DCL can be called an Analog Differential Clipper (ADiC)~\cite{Nikitin19ADiCpatentCIPs}.

\subsection{Basic ADiC structure} \label{subsec:basic ADiC}
\begin{figure}[!b]
\centering{\includegraphics[width=8.6cm]{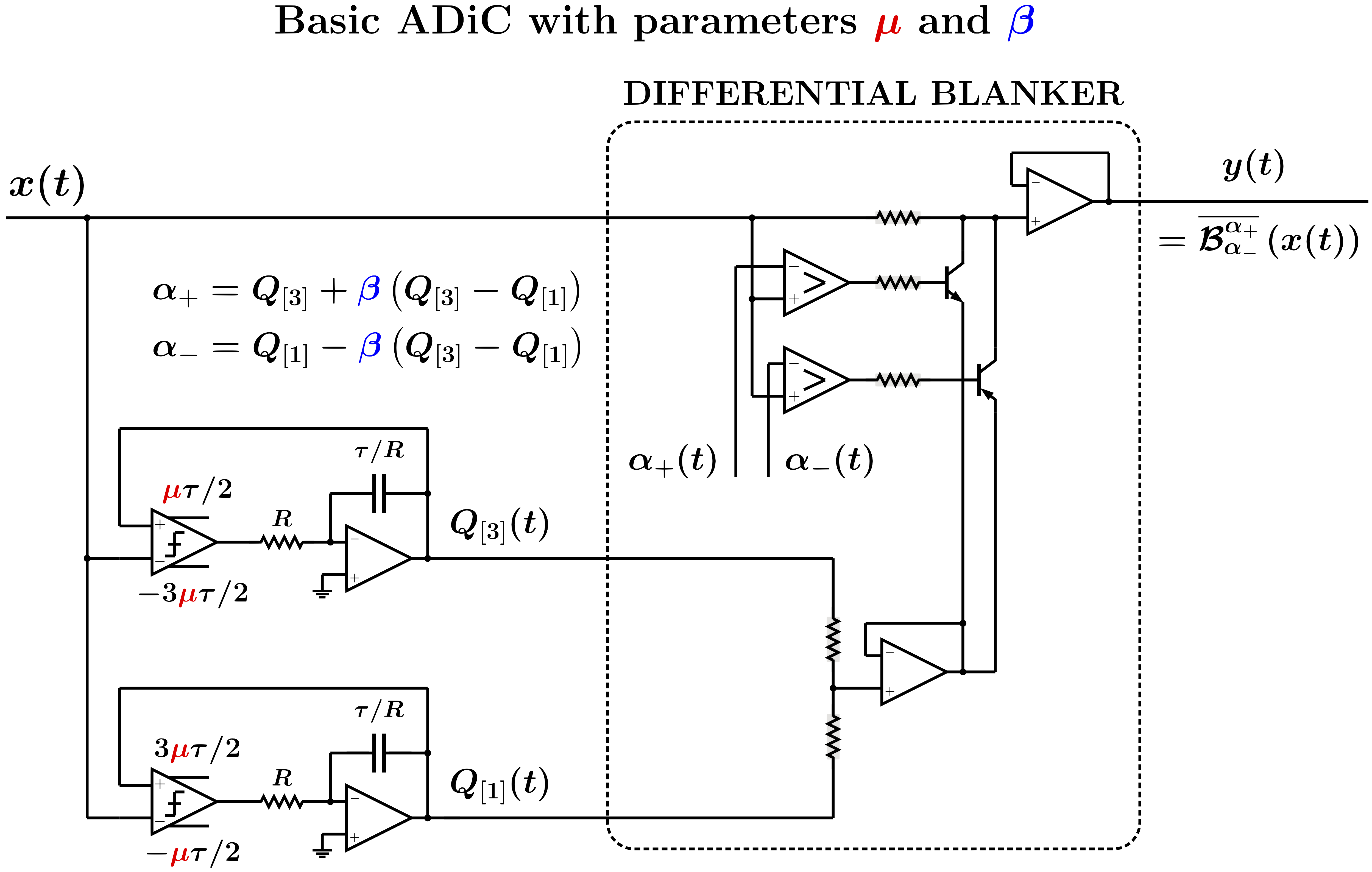}}
\caption{Simplified schematic of basic ADiC.
\label{fig:ADiC schematic}}
\end{figure}

\begin{figure}[!b]
\centering{\includegraphics[width=8.6cm]{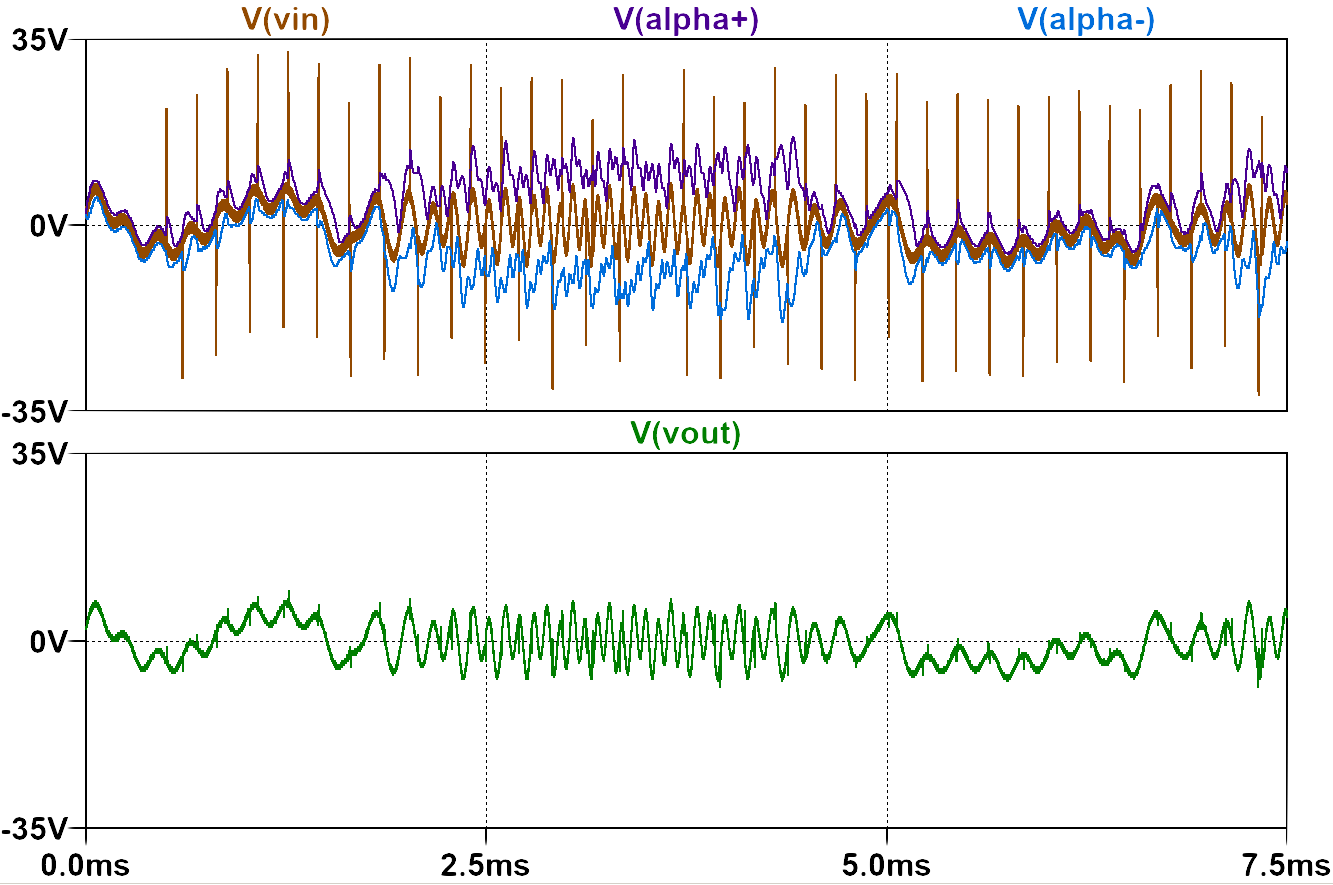}}
\caption{Illustrative traces from LTspice simulation of basic ADiC circuit.
\label{fig:ADiC LTspice}}
\end{figure}

Figs.\,\ref{fig:basic ADiC} and~\ref{fig:ADiC schematic} show a block diagram and a simplified schematic, respectively, of a basic ADiC structure. In this simple ADiC the range is constructed by linear combinations of the outputs of QTFs for the 1st and the 3rd quartiles, and the mid-range is the arithmetic mean of these QTF outputs. Fig.\,\ref{fig:ADiC LTspice} provides an illustration of the signal traces for the input, the output, and the fences from an LTspice simulation of the simple circuit shown in Fig.\,\ref{fig:ADiC schematic}.

\subsection{Feedback-based ADiC} \label{subsec:ADiC}
\begin{figure}[!t]
\centering{\includegraphics[width=8.6cm]{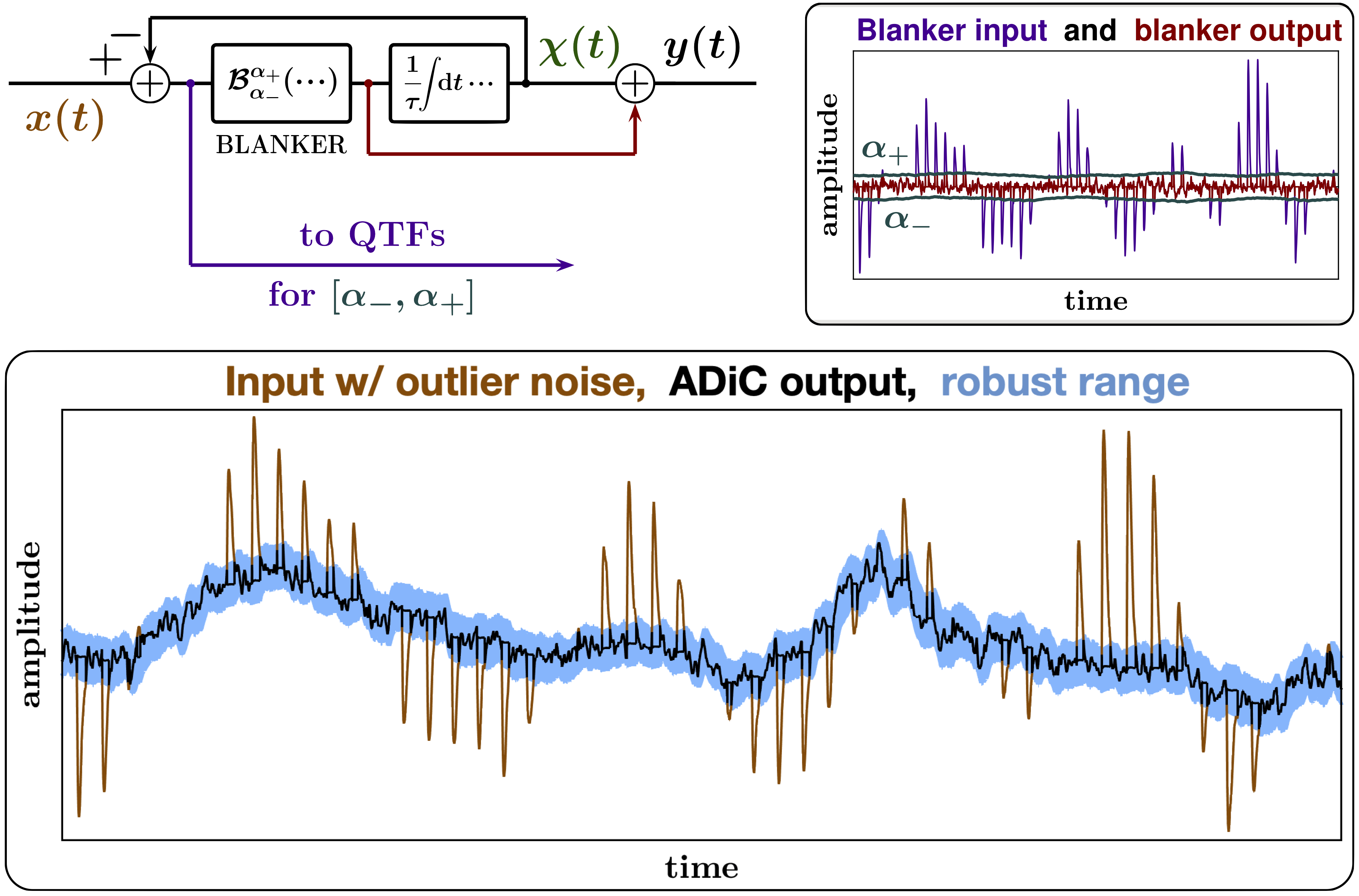}}
\caption{Feedback-based ADiC replacing outliers with~$\chi(t)$.
\label{fig:ADiC}}
\end{figure}

Fig.\,\ref{fig:ADiC} presents a feedback-based ADiC variant that has a number of practical advantages and is well suited for mitigation of hidden outlier noise~\cite{Nikitin19ADiCpatent, Nikitin18ADiC-ICC}. As the diagram in the upper left of the figure shows, the ADiC output~$y(t)$ can be described as
\beginlabel{equation}{eq:ADiC equation}
  \left\{
  \begin{array}{cc}
    \!\! y(t) = \chi(t) + \tau\dot{\chi}(t)\\[1mm]
    \!\! \displaystyle \dot{\chi}(t) = \frac{1}{\tau}\, \BalphaPM\left(x(t)\!-\!\chi(t) \right)
  \end{array}\right.,
\end{equation}
where~$x(t)$ is the input signal, $\chi(t)$~is the DCL, the {\it blanking function\/}~$\BalphaPM(x)$ is a particular type of an {\it influence function}~\cite{Hampel74influence} that is defined as
\beginlabel{equation}{eq:clipping function}
  \BalphaPM(x)  = \left\{
  \begin{array}{cc}
    \!\! x & \mbox{for} \quad \alpha_- \le x \le \alpha_+\\
    \!\! 0 & \mbox{otherwise}
  \end{array}\right.,
\end{equation}
and where~$[\alpha_-,\alpha_+]$ is a robust range for the {\it difference signal\/}~${x(t)-\chi(t)}$ (the {\it blanking range\/}). Thus an ADiC is an {\it intermittently nonlinear\/} filter that outputs the DCL~$\chi(t)$ only when outliers in the difference signal are detected, performing outlier noise mitigation without modifying the input signal otherwise. For the range fences such that ${\alpha_- \le x(t)\!-\!\chi(t) \le \alpha_+}$ for all~$t$, the DCL~$\chi(t)$ is the output of a 1st~order linear lowpass filter with the 3\,dB corner frequency~$1/(2\pi\tau)$. However, when an outlier of the difference signal is encountered, the rate of change of~${\chi(t)}$ is zero and the DCL maintains its previous value for the duration of the outlier.


\subsection{ADiC-based outlier noise filtering} \label{subsec:ADiC-based filtering}
As noted in Section~\ref{subsec:ADiC}, in the absence of outlier noise the difference signal of the feedback-based ADiC is the output of a 1st~order highpass filter with the 3\,dB cutoff frequency~$1/(2\pi\tau)$. Consequently, as follows from the discussion in Section~\ref{sec:elusive nature}, for efficient outlier noise mitigation the ADiC's {\it time parameter\/}~$\tau$ should be sufficiently large so that such a filter does not significantly reduce the bandwidth of the noise. On the other hand, unless the amplitude of the signal of interest is much smaller than a typical amplitude of the noise outliers, $\tau$~should be sufficiently small so that such 1st~order highpass filter noticeably reduces the amplitude of the signal of interest. In other words, $\tau$~should be sufficiently small so that a 1st~order lowpass filter with the corner frequency~$1/(2\pi\tau)$ does not significantly affect the signal of interest. Such a compromise is much easier to reach for a low-frequency signal of interest, but is more challenging to achieve for a passband signal. Therefore, the best application for an ADiC would be the removal of outliers from the ``excess band" noise (see Section~\ref{subsec:excess band}), when the signal of interest is mainly excluded. Then ADiC-based filtering that mitigates wideband outlier noise while preserving the signal of interest can be accomplished as described below.

\begin{figure}[!b]
\centering{\includegraphics[width=8.6cm]{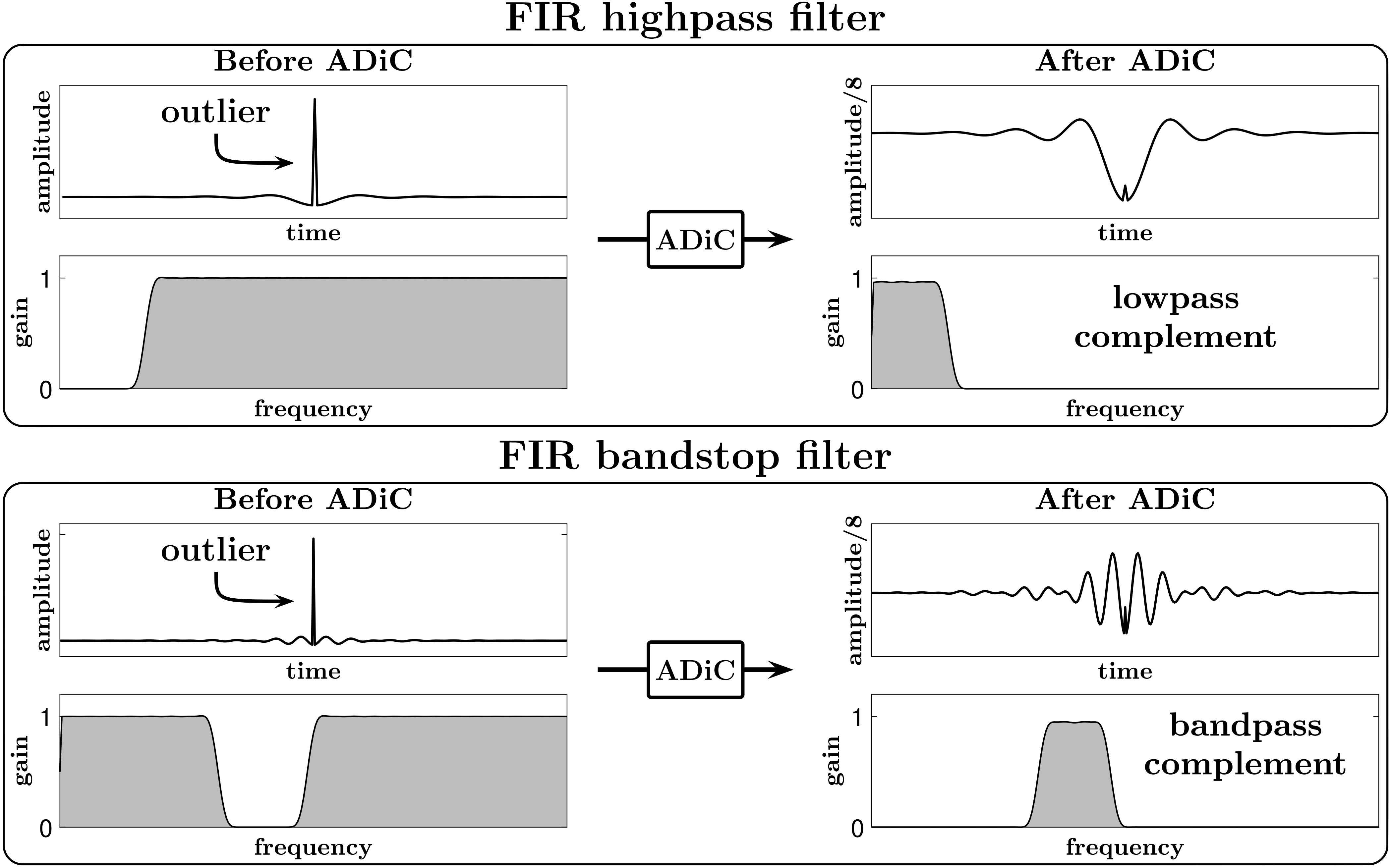}}
\caption{Spectral inversion by ADiC.
\label{fig:inversion}}
\end{figure}
\begin{figure}[!b]
\centering{\includegraphics[width=8.6cm]{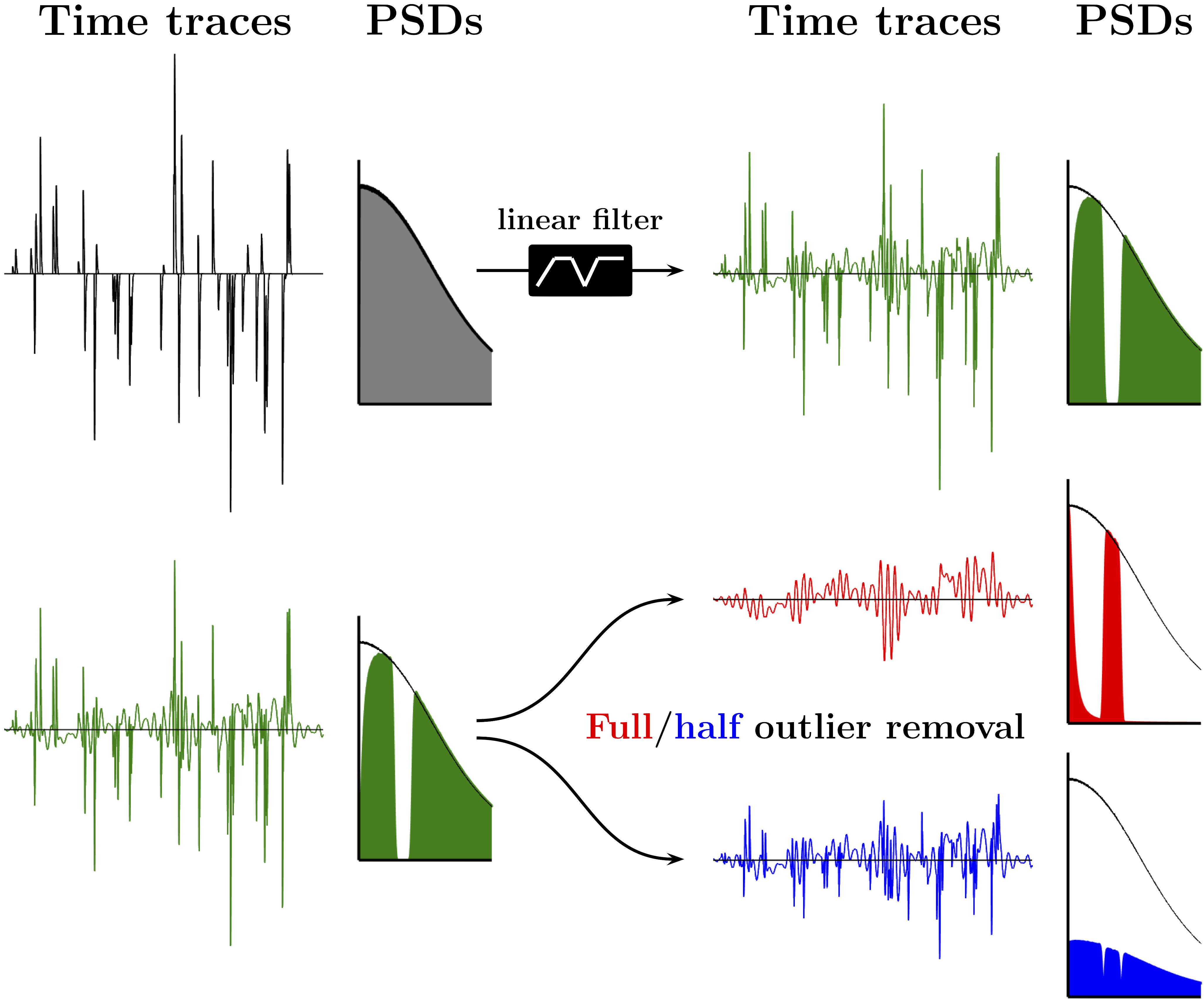}}
\caption{Spectral ``cockroach effect" caused by outlier removal.
\label{fig:reshaping}}
\end{figure}
\begin{figure}[!b]
\centering{\includegraphics[width=8.6cm]{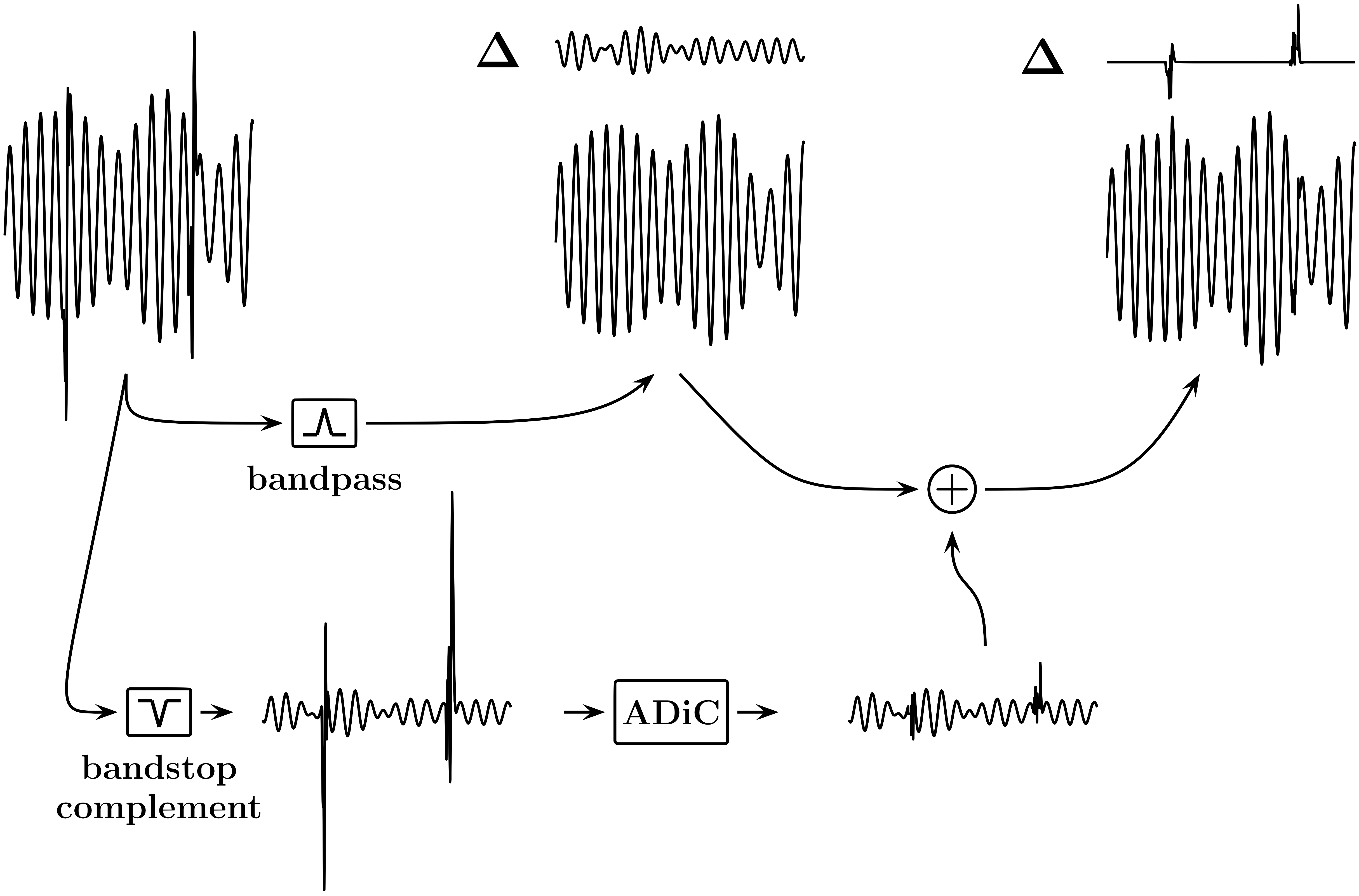}}
\caption{ADiC-based outlier noise mitigation for passband signal.
\label{fig:passband mitigation}}
\end{figure}
\begin{figure*}[!t]
\centering{\includegraphics[width=16.4cm]{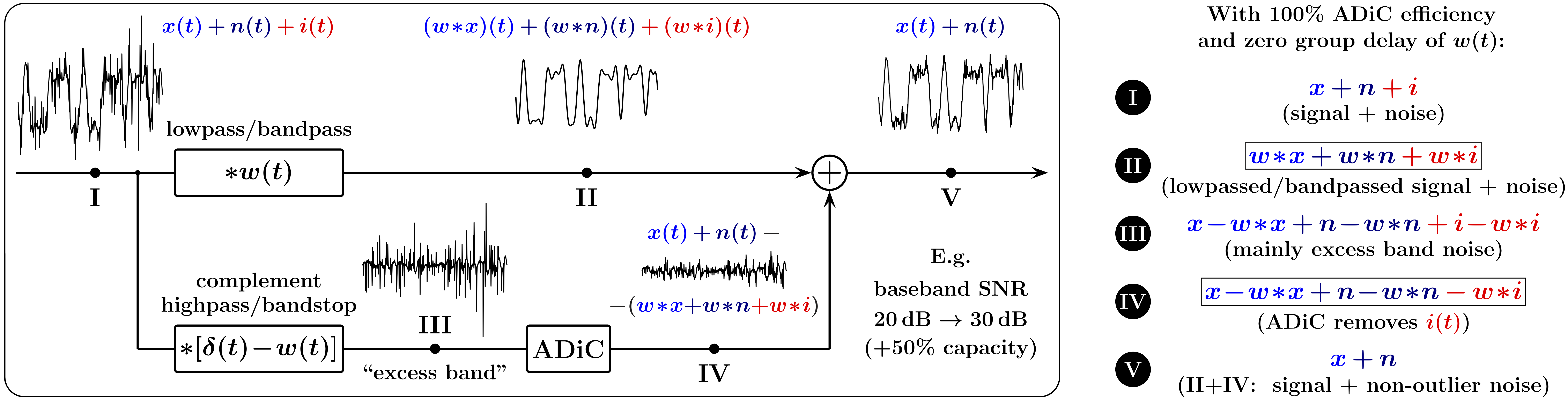}}
\caption{Complementary ADiC filter (CAF) for removing wideband noise outliers while preserving band-limited signal of interest.
\label{fig:full ADiC}}
\end{figure*}

\subsubsection{Spectral inversion by ADiC and {\em ``efecto cucaracha"\/}} \label{subsubsec:cucaracha}
Let us first note that applying an ADiC to an impulse response of a highpass and/or bandstop filter containing a distinct outlier causes the ``spectral inversion" of the filter, transforming it into its {\it complement\/}, e.g. a highpass filter into a lowpass, and a bandstop filter into a bandpass filter. This is illustrated in Fig.~\ref{fig:inversion} where, for simplicity, FIR filters are used. Hence, as further demonstrated in Fig.~\ref{fig:reshaping}, an ADiC applied to a filtered outlier noise can significantly reshape its spectrum. Such spectral reshaping by an ADiC can be called {\it ``efecto cucaracha"\/} (``cockroach effect"), when reducing the effects of outlier noise in some spectral bands increases its PSD in the bands with previously low outlier noise PSD. We can use this property of an ADiC for removing outlier noise while preserving the signal of interest, and for addressing complex interference scenarios.

\subsubsection{Removing outlier noise while preserving signal of interest} \label{subsubsec:full ADiC}
For example, in Fig.~\ref{fig:passband mitigation} the bandpass filter mainly matches the signal's passband, and the bandstop filter is its ``complement" obtained by spectral inversion of the bandpass filter, so that the sum of the outputs of the bandpass and the bandstop filters is equal to the input signal. The input passband signal of interest affected by a wideband outlier noise can be seen in the upper left of Fig.~\ref{fig:passband mitigation}. The output of the bandpass filter is shown in the upper middle of the figure, where the trace marked by ``$\Delta$" shows the effect of the outlier noise on the passband signal. As discussed in Section~\ref{subsec:excess band}, the output of the bandstop filter is mainly the ``excess band" noise. After the outliers of the excess band noise are mitigated by an ADiC, the remaining excess band noise is added to the output of the bandpass filter. As the result, the combined output (seen in the upper right of the figure) will be equal to the original signal of interest affected by a wideband noise with reduced outliers. This mitigated outlier noise is shown by the trace marked by ``$\Delta$" in the upper right.

Fig.~\ref{fig:full ADiC} summarizes such ``complementary" ADiC-based outlier noise removal from band-limited signals. To simplify the mathematical expressions, we here use zero for the group delay of the linear filters and assume that the ADiC completely removes the outlier component~$i(t)$ from the excess band. We will call this ADiC-based filtering structure a Complementary ADiC Filter (CAF).

\subsubsection{ADiC vs. linear: Effect on channel capacity} \label{subsubsec:capacity}
\begin{figure}[!b]
\centering{\includegraphics[width=8.6cm]{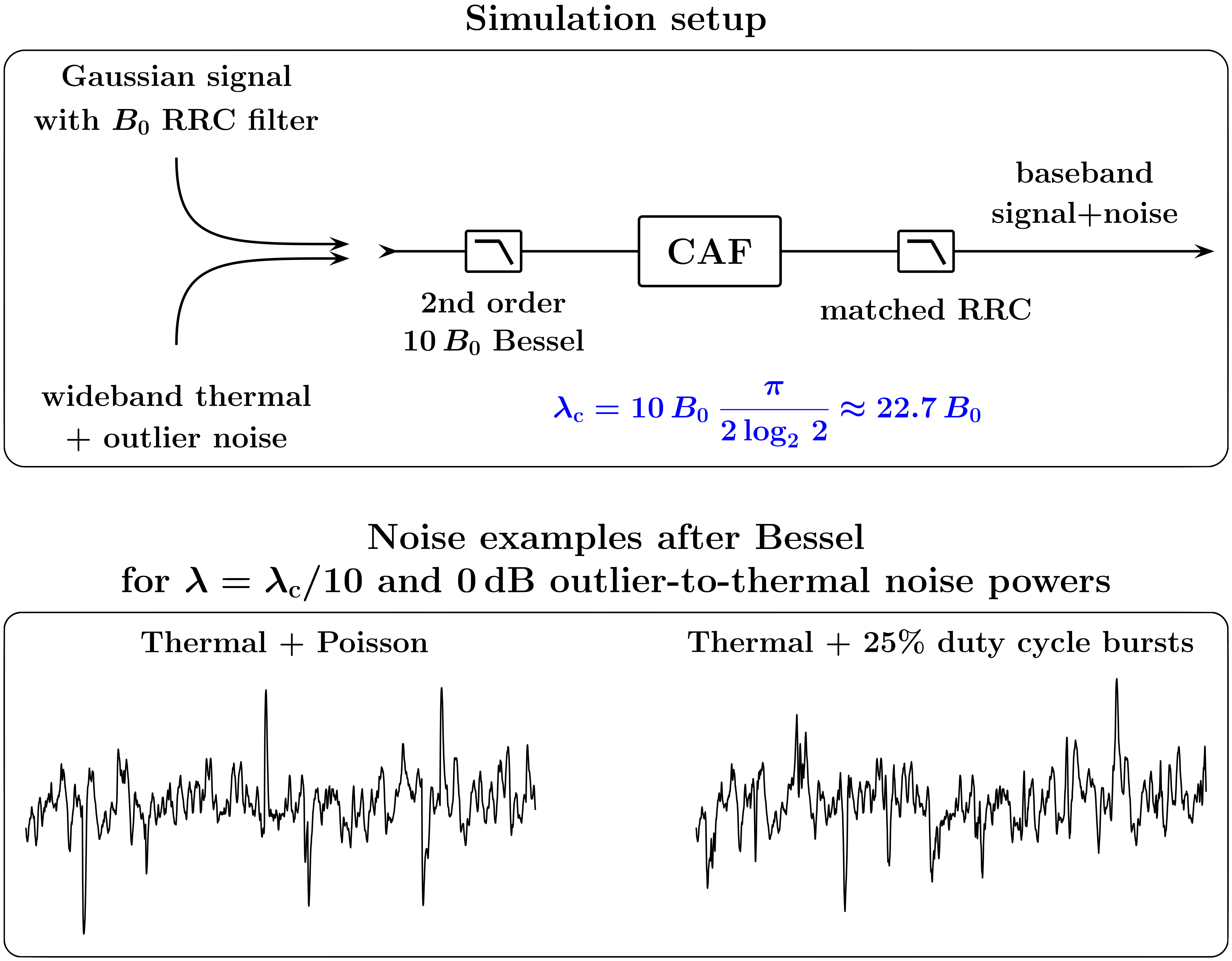}}
\caption{Simulation setup and noise examples.
\label{fig:simulation setup}}
\end{figure}

Fig.~\ref{fig:simulation setup} outlines the simulation setup for quantification of the potential improvements in signal quality provided by ADiC-based mitigation of outlier noise in comparison with linear filtering. The signal of interest is formed as white Gaussian noise filtered with a root-raised-cosine (RRC) filter with the nominal bandwidth~$B_0$, and a mixture of wideband thermal and outlier noise is added to the signal. The front-end filter is a 2nd~order lowpass Bessel with the cutoff frequency~$10B_0$. The time-bandwidth product of a lowpass Bessel filter is approximately that of a Gaussian filter, ${2\log_2(2)/\pi}$, and thus~$\lambda_{\rm c}\approx 22.7B_0$ is the ``pileup threshold" rate of the front-end filter. As discussed in Section~\ref{subsec:what hides}, for outlier arrival rates significantly above~$\lambda_{\rm c}$ the outlier noise becomes effectively Gaussian and can no longer be efficiently mitigated. The lower panel of Fig.~\ref{fig:simulation setup} provides noise examples at the output of the front-end filter, for Poisson noise with normally distributed amplitudes, and for periodic Gaussian bursts with 25\% duty cycle. The ADiC-based filter with the topology shown in Fig.~\ref{fig:full ADiC} (CAF) processes the front-end filter output, and the baseband signal is obtained by applying the matched RRC filter to the CAF output.

\begin{figure}[!t]
\centering{\includegraphics[width=8.6cm]{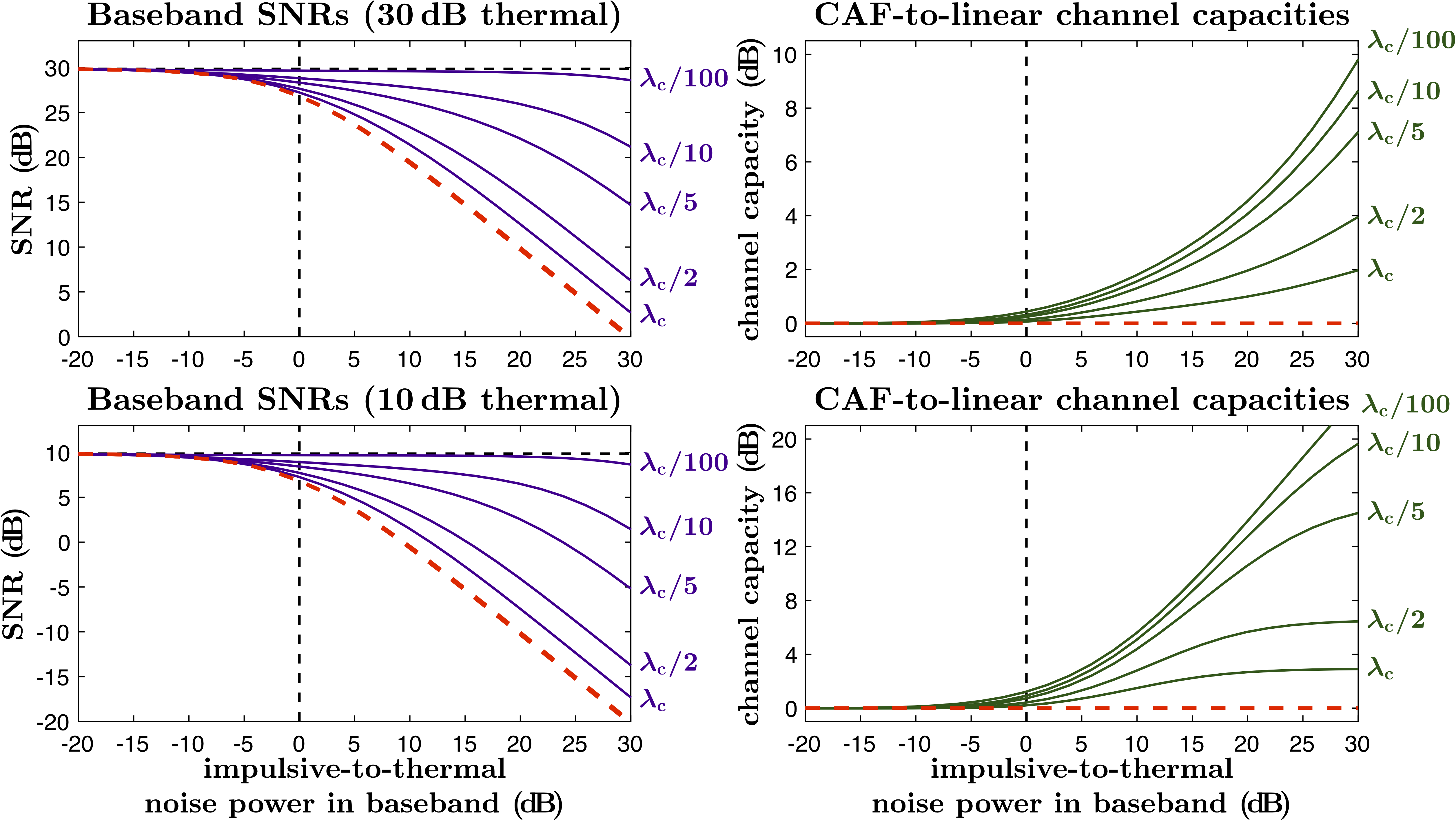}}
\caption{Poisson noise with normally distributed amplitudes.
\label{fig:Poisson}}
\end{figure}

\begin{figure}[!t]
\centering{\includegraphics[width=8.6cm]{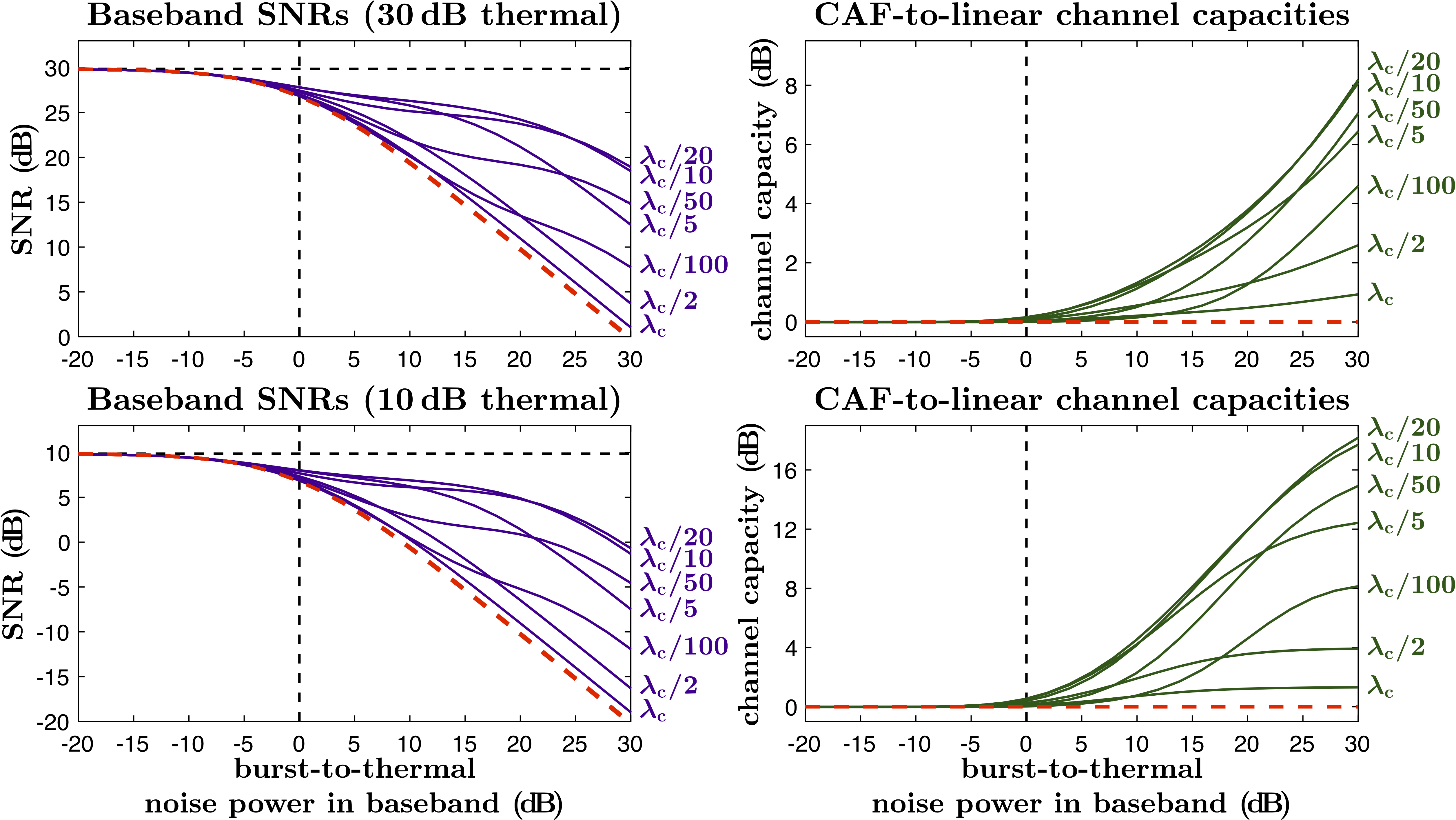}}
\caption{Periodic Gaussian bursts with 10\% duty cycle.
\label{fig:bursts rates}} 
\end{figure}

\begin{figure}[!t]
\centering{\includegraphics[width=8.6cm]{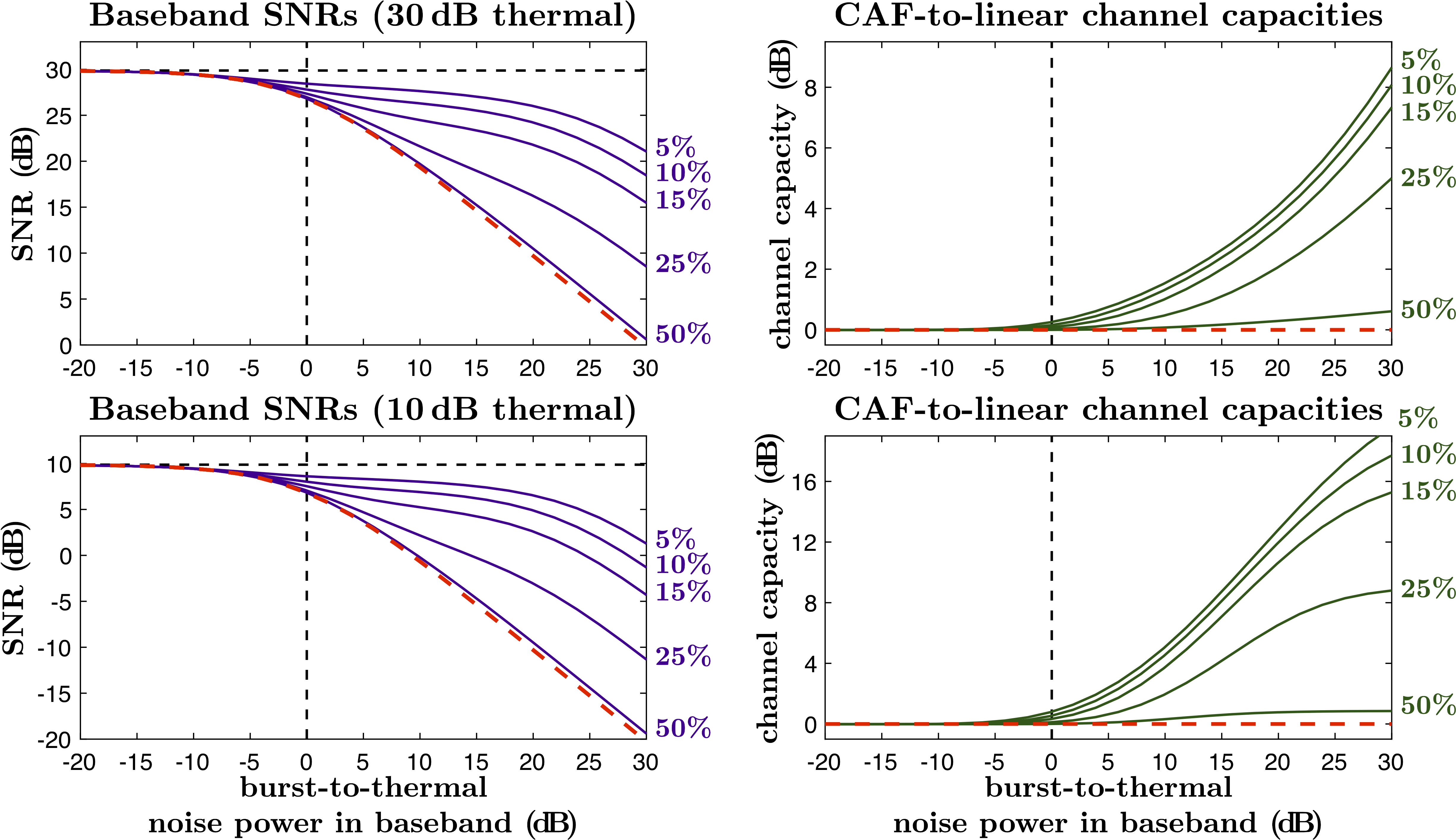}}
\caption{$\!$Periodic Gaussian bursts with $\lambda\!=\!\lambda_{\rm c}/10$ and different duty cycles.
\label{fig:bursts DC}}
\end{figure}

Figs.~\ref{fig:Poisson}, \ref{fig:bursts rates}~and~\ref{fig:bursts DC} illustrate the improvements in the baseband SNRs and in the channel capacities, as functions of the outlier-to-thermal noise power in the baseband, for different outlier noise compositions and moderate (10\,dB) and high (30\,dB) thermal noise SNRs. Since ADiC-based filtering removes noise outliers, the baseband noise after such filtering is effectively Gaussian, and the Shannon formula~\cite{Shannon49communication} can be used to calculate the limit on the channel capacity. However, the baseband noise after linear filtering (without CAF in the signal chain) may not be Gaussian, especially for low rates and high outlier noise powers. Nevertheless, we still use the Shannon formula as a proxy measure for the capacity of the linear channel, to quantify the comparative signal quality improvement.

In all these simulations, a ``default" set of CAF parameters was used, with the ``no harm" constraint such that nonlinear filtering does not degrade the resulting signal quality, as compared with the linear filtering, for any signal+noise mixtures. Thus, while providing resistance to outlier noise, in the absence of such noise CAFs behave effectively linearly, avoiding the detrimental effects, such as distortions and instabilities, often associated with nonlinear filtering. The ``no harm" property is especially important when considering complex, highly nonstationary interference scenarios, e.g.  in mobile and cognitive communication systems where the transmitter positions, powers, signal waveforms, and/or spectrum allocations vary dynamically. Note that when a CAF improves the signal quality, its performance can be further enhanced by optimizing its parameters.


\subsection{Analog vs. digital implementations} \label{subsec:analog vs digital}
The concept of ADiC-based filtering relies on continuous-time (analog) operations such as differentiation, antidifferentiation, and analog convolution. Therefore the most natural platform for implementing such filtering is analog circuitry. Analog processing is very appealing, e.g., when the requirements include inherently real-time operation, higher bandwidth, and lower power. On the other hand, digital processing offers simplified development and testing, configurability, and reproducibility. In addition, different ADiC components (e.g. QTFs) can be easily included into numerical algorithms without a need for separate circuits, and digital ADiC-based filtering is simpler to extend to complex-valued processing and to incorporate into various machine learning and optimization-based approaches.

\subsubsection{Digital: Where to get bandwidth?} \label{subsubsec:bandwidth}
\begin{figure}[!t]
\centering{\includegraphics[width=8.6cm]{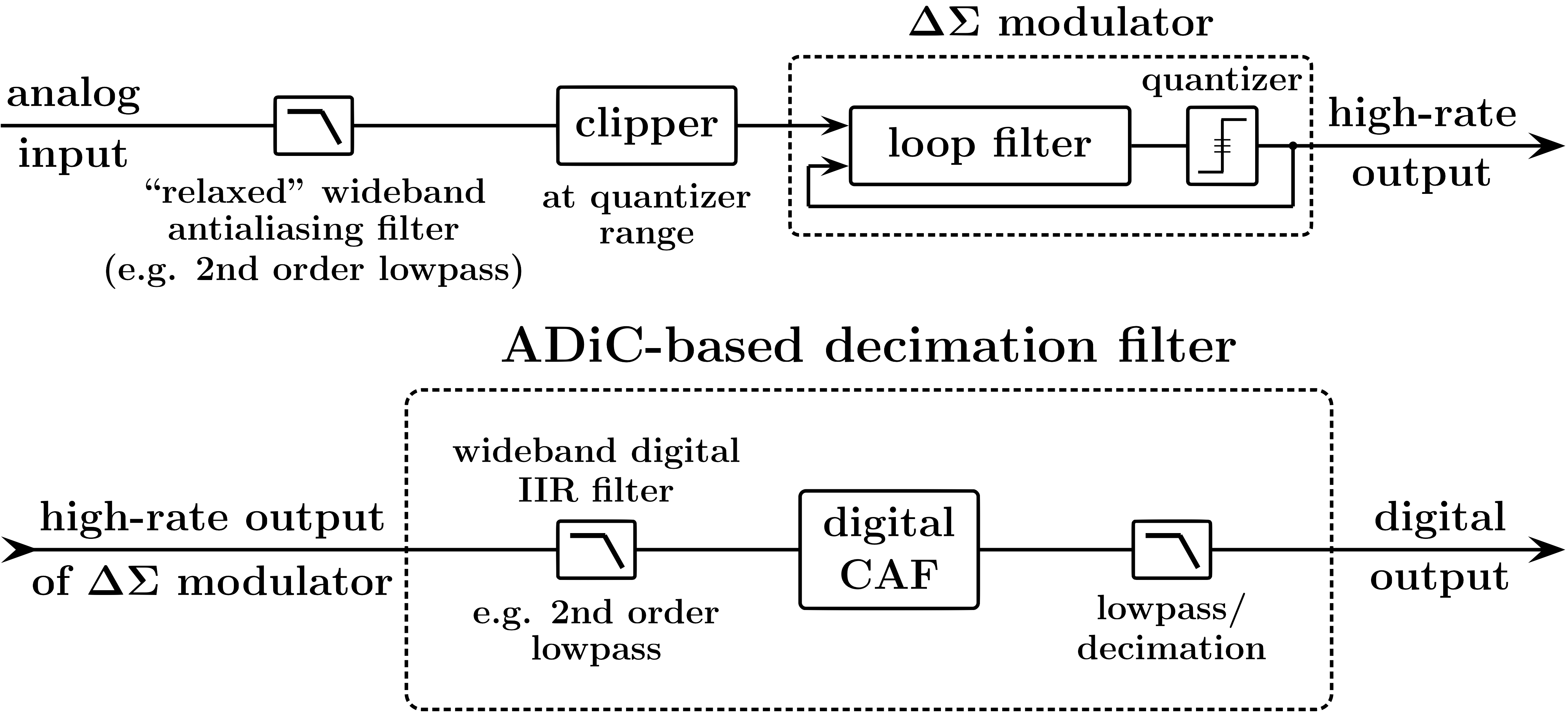}}
\caption{ADiC filtering in digital domain following $\Delta\Sigma$ modulator.
\label{fig:DeltaSigma}}
\end{figure}

As discussed in Section~\ref{sec:elusive nature}, efficient mitigation of wideband outlier noise requires availability of a sufficiently broad excess band and the respectively high ADC sampling rate. In addition, the sampling rate needs to be further increased so that analog differentiation can be replaced by its accurate finite-difference approximation to enable ``effectively analog" processing. Fig.~\ref{fig:DeltaSigma} illustrates how inherently high oversampling rate of a $\Delta\Sigma$~ADC can be used to trade amplitude resolution for higher sampling rate and thus to enable such efficient digital ADiC-based filtering.
Presently, $\Delta\Sigma$ ADCs are used for converting analog signals over a wide range of frequencies, from DC to several megahertz. These converters comprise a highly oversampling modulator followed by a digital/decimation filter that together produce a high-resolution digital output~\cite{Bourdopoulos03Delta-Sigma, DataConversionHandbook, Geerts06design}.

\begin{figure}[!b]
\centering{\includegraphics[width=8.6cm]{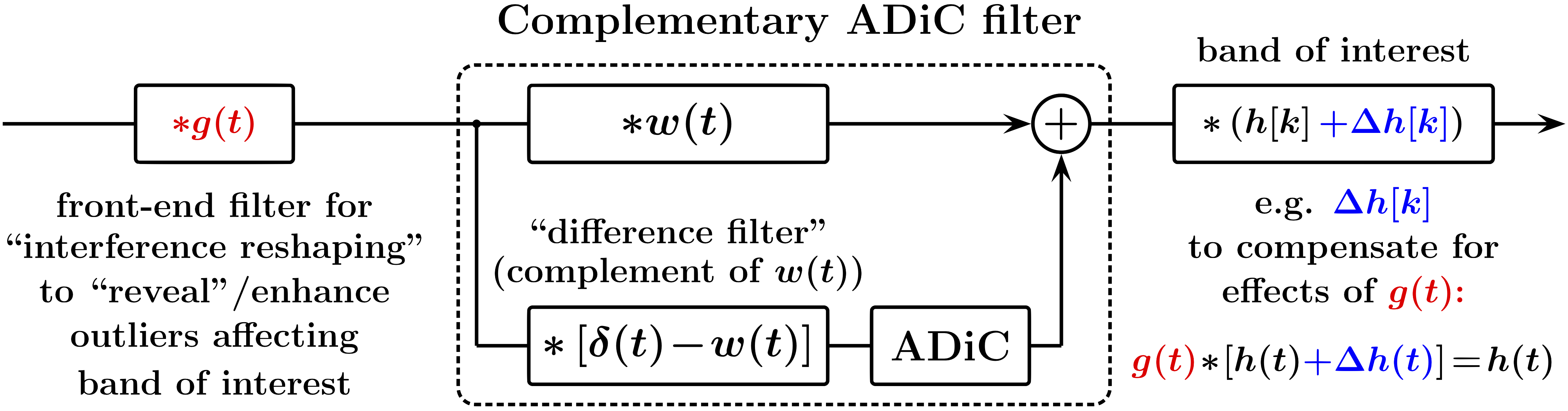}}
\caption{Addressing complex interference scenarios.
\label{fig:complex}}
\end{figure}

The high sampling rate allows the use of ``relaxed," wideband antialiasing filters to ensure the availability of sufficiently wide excess band. As a practical matter, wideband filters with a flat group delay and a small time-bandwidth product (e.g. with a Bessel response) should be used in order to increase the mitigable rates. Further, a simple clipper is employed ahead of the modulator to limit the magnitude of excessively strong outliers in the input signal, thus preventing the modulator from saturation. The low (e.g. 1-bit) amplitude resolution of the output of the $\Delta\Sigma$ modulator does not allow direct application of a digital ADiC. However, since the oversampling rate is significantly higher (e.g. by two to three orders of magnitude) than the Nyquist rate of the signal of interest, a wideband digital filter can be first applied to the output of the quantizer to enable the ADiC-based outlier filtering. To reduce computations and memory requirements, such a filter can be an infinite impulse response (IIR) filter. For instance, for a 1-bit $\Delta\Sigma$ modulator with a 20\,MHz clock, and a required 100\,kS/s decimated output, the bandwidth of the wideband IIR filter ahead of the CAF in Fig.~\ref{fig:DeltaSigma} can be about 500\,kHz. Furthermore, the analog antialiasing filter and the wideband IIR filter should be co-designed to ensure the desired excess band response (in both time- and frequency domains). For example, the corner frequencies and the quality factors of the 2nd~order analog antialiasing and the wideband digital IIR lowpass filters shown in Fig.~\ref{fig:DeltaSigma} can be chosen to ensure that the combined response of these cascaded filters is that of the 4th~order Bessel-Thomson filter~\cite{Schaumann01DesignOfAnalogFilters, Proakis06digital}.

\subsubsection{Addressing complex interference scenarios} \label{subsubsec:complex}
The temporal and/or amplitude structures (and thus the distributions) of non-Gaussian signals are generally modifiable by linear filtering (e.g. see Fig.~\ref{fig:origins}), and non-Gaussian interference can often be converted from sub-Gaussian into super-Gaussian, and {\it vice versa\/}, by such filtering~\cite{Nikitin13adaptive, Nikitin15OOB}. Therefore the ability of CAFs to mitigate impulsive (super-Gaussian) noise can translate into mitigation of non-Gaussian noise and interference in general, including sub-Gaussian noise (e.g. wind noise at microphones). For example, as illustrated in Fig.~\ref{fig:complex}, a linear filter can be employed ahead of the CAF to enhance the outliers affecting the band of interest and perform analog-to-digital conversion combined with mitigation of this interference. Subsequently, if needed, the digital decimation filter can be modified to compensate for the impact of the front-end filter on the signal of interest.

\begin{figure}[!t]
\centering{\includegraphics[width=8.6cm]{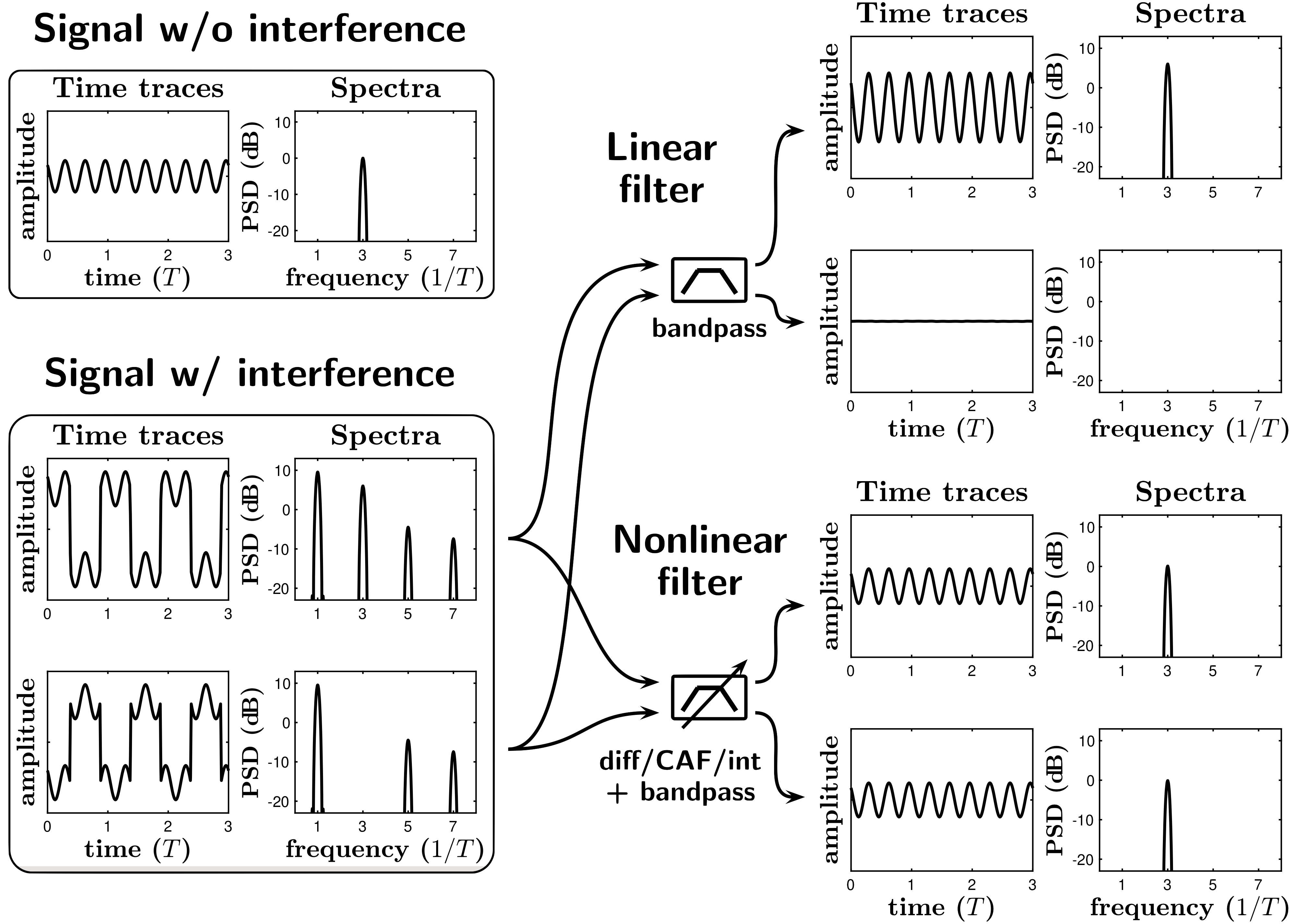}}
\caption{Toy example of suppressing square wave interference.\label{fig:toy example2}}
\end{figure}
\begin{figure}[!b]
\centering{\includegraphics[width=8.6cm]{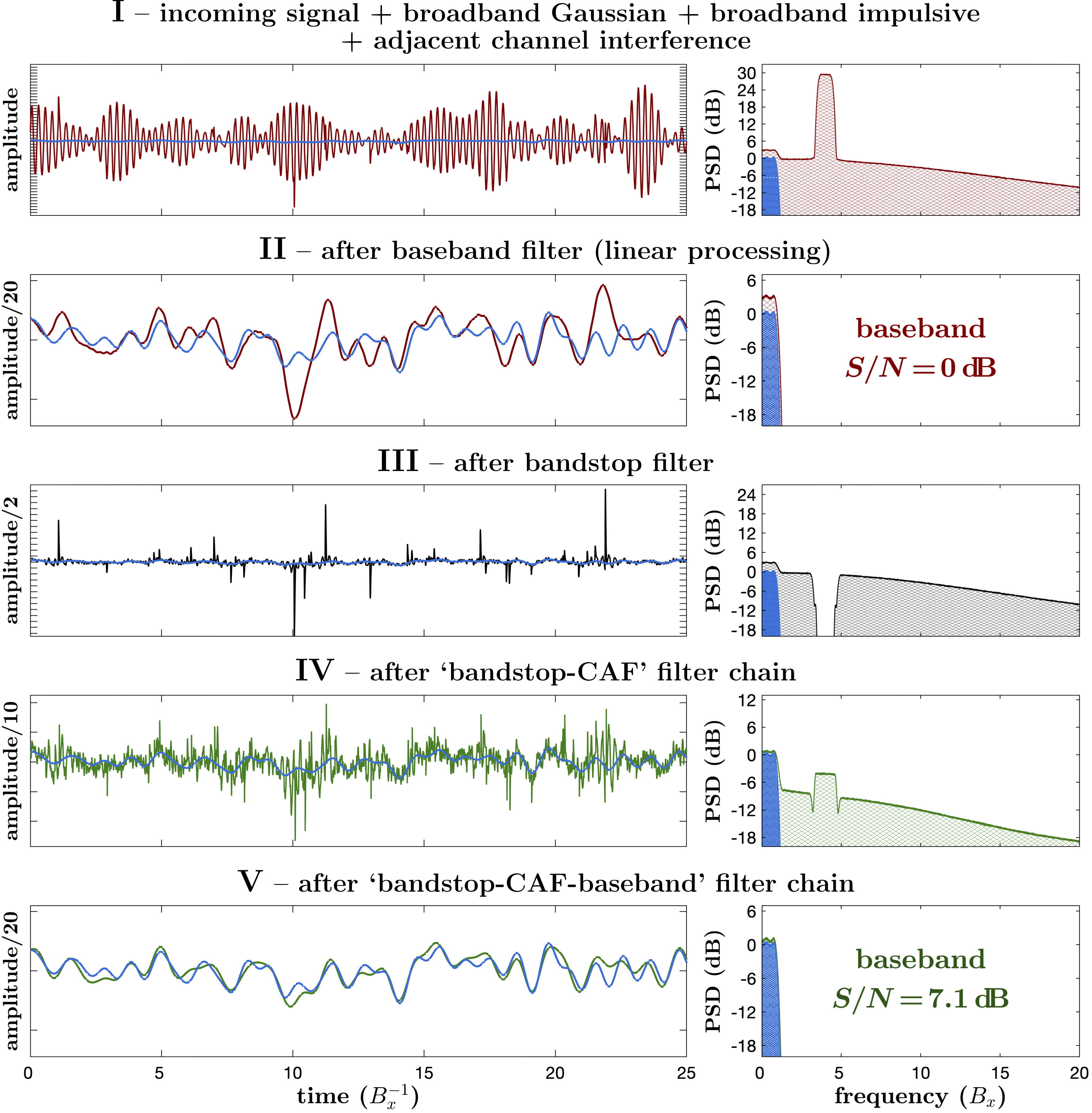}}
\caption{CAF vs. linear for strong adjacent channel interference.
\label{fig:adjacent traces}}
\end{figure}
\begin{figure}[!b]
\centering{\includegraphics[width=8.6cm]{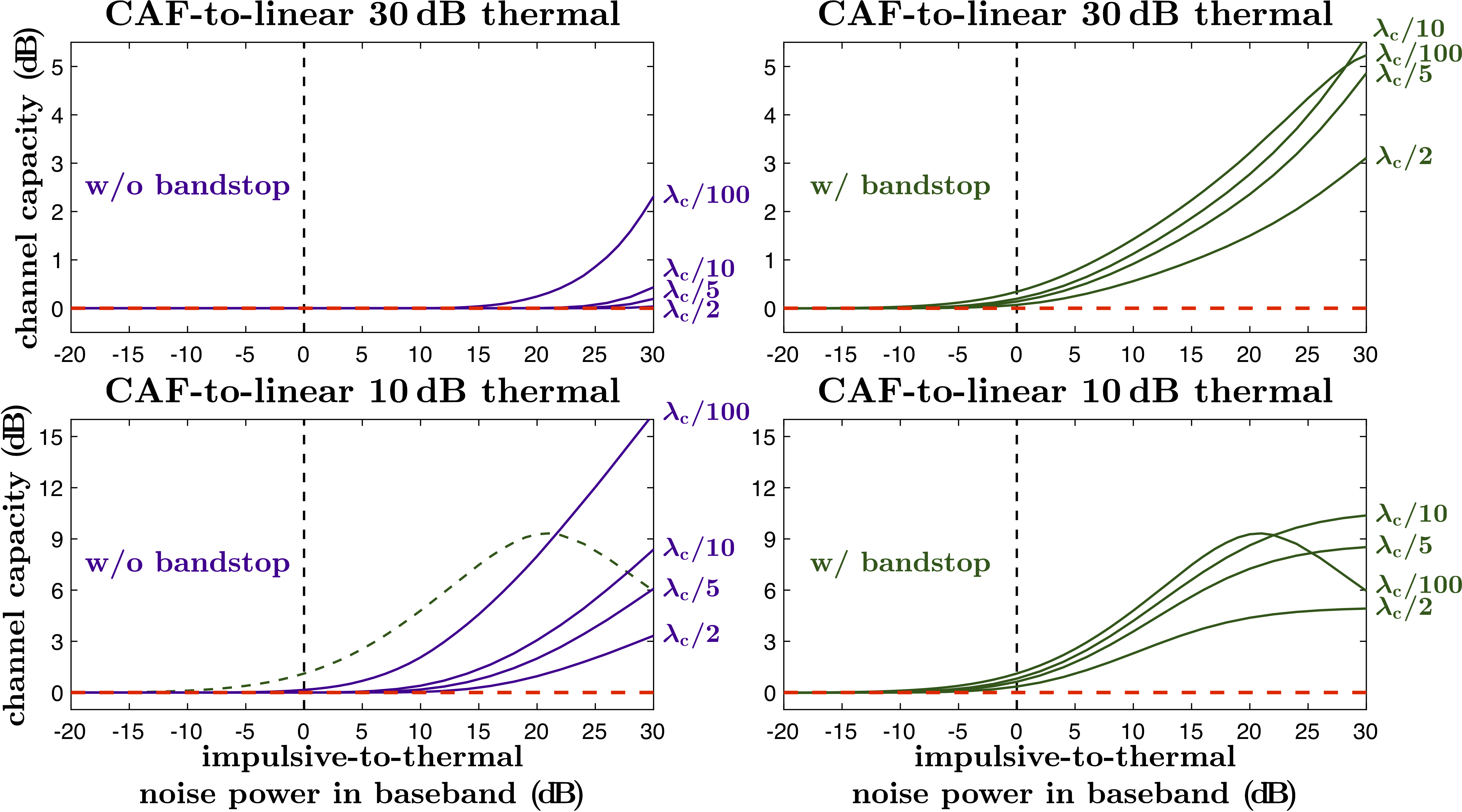}}
\caption{CAF vs. linear with and without front-end bandstop.
\label{fig:adjacent capacities}}
\end{figure}

The toy example of Fig.~\ref{fig:toy example2} illustrates this approach for a mixture of a sine wave with the period~$T/3$ (shown in the upper left) and a square wave with the period~$T$. As can be seen in the lower left of the figure, the 3rd~harmonic of the square wave interferes with the sine wave either constructively or destructively, and the power of this harmonic is equal to that of the signal. A linear bandpass filter can neither reduce the power of the constructive interference nor restore the ``missing" signal in the case of the destructive interference. This is shown in the upper right of the figure. Although a square wave is a sub-Gaussian signal with a negative peakedness ($-4.77\,$dBG), its time derivative is a super-Gaussian impulse train that can be efficiently mitigated by an ADiC or a CAF. On the other hand, a time derivative of a sine wave is still a sine wave. Hence, as illustrated in the lower right of Fig.~\ref{fig:toy example2}, applying a CAF to a derivative of the sine+square wave mixture and integrating the CAF output before bandpass filtering effectively suppresses the square wave interference.

Fig.~\ref{fig:adjacent traces} provides a practical example of using a front-end filter to enhance the performance of a CAF in the presence of strong adjacent channel interference. Such interference obscures the wideband impulsive noise (see panel~I), making CAF ineffective. A bandstop filter suppressing the adjacent channel interference ``reveals" the impulsive noise affecting the baseband (panel~III), enabling its efficient mitigation by a CAF. Note that, as can be seen in panel~IV, due to the ``cockroach effect" CAF increases the PSD of the impulsive noise in the stopband of the bandstop filter. However, this does not affect the baseband SNR as the baseband filter suppresses the noise outside of the baseband.

For the strong adjacent channel interference shown in Fig.~\ref{fig:adjacent traces} (i.e. with the PSD $30\,$dB larger than that of the signal of interest), Fig.~\ref{fig:adjacent capacities} further illustrates the improvements in the channel capacities, as functions of the impulsive-to-thermal noise power in the baseband, for different impulsive noise rates with moderate (10\,dB) and high (30\,dB) thermal noise SNRs, and with and without the front-end bandstop filter. In this example, the same setup and default set of CAF parameters was used as in the simulations of Figs.~\ref{fig:Poisson}, \ref{fig:bursts rates}~and~\ref{fig:bursts DC}, with the ``no harm" constraint such that nonlinear filtering does not degrade the resulting signal quality for any signal+noise mixtures. Note that the bandstop filter significantly increases the effectiveness of the impulsive noise suppression by a CAF in the presence of adjacent channel interference, and more noticeably for higher thermal noise SNRs.

\begin{figure}[!b]
\centering{\includegraphics[width=8.6cm]{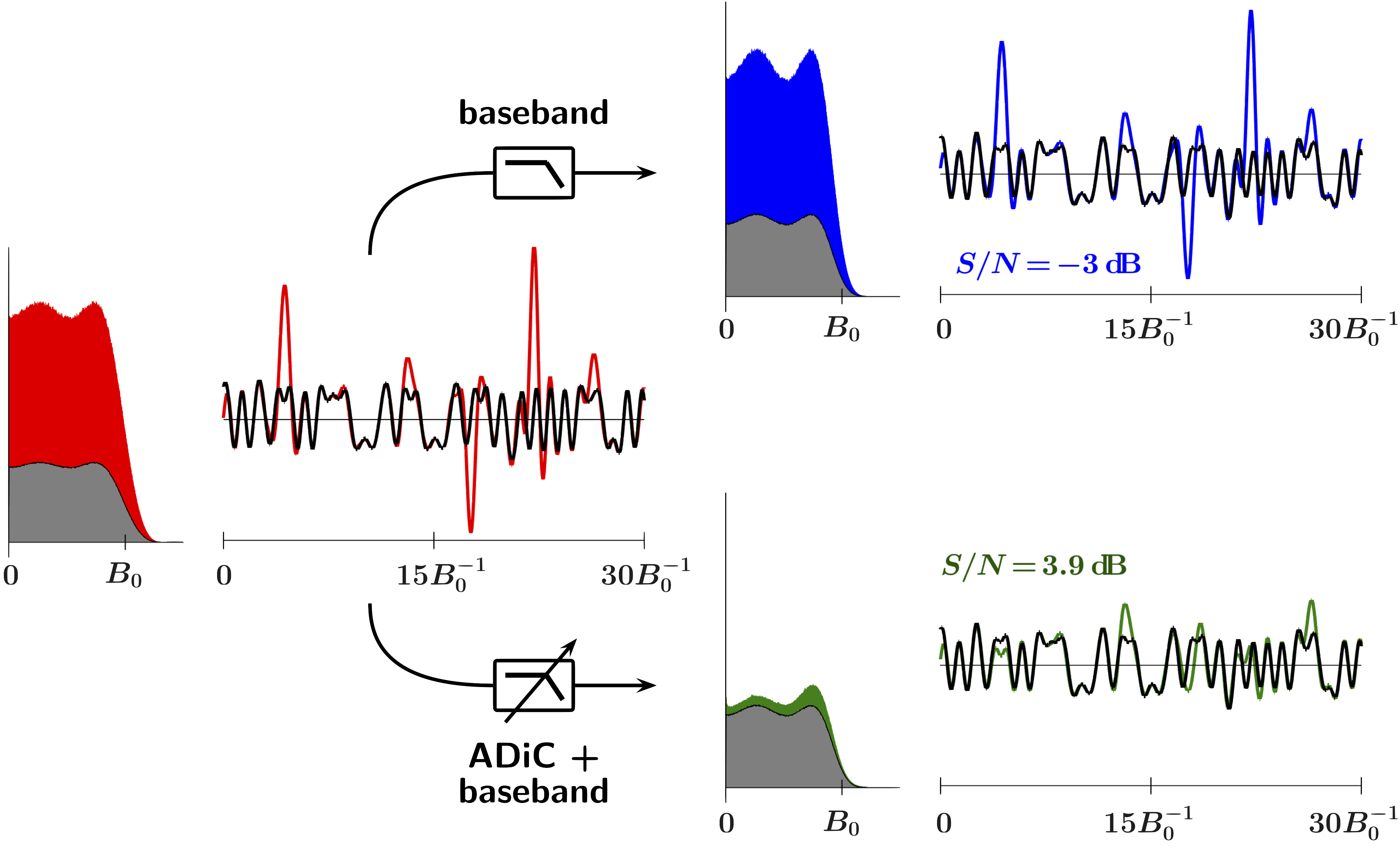}}
\caption{ADiC vs. linear in ``shared band" case.
\label{fig:shared band traces}}
\end{figure}
\begin{figure}[!b]
\centering{\includegraphics[width=8.6cm]{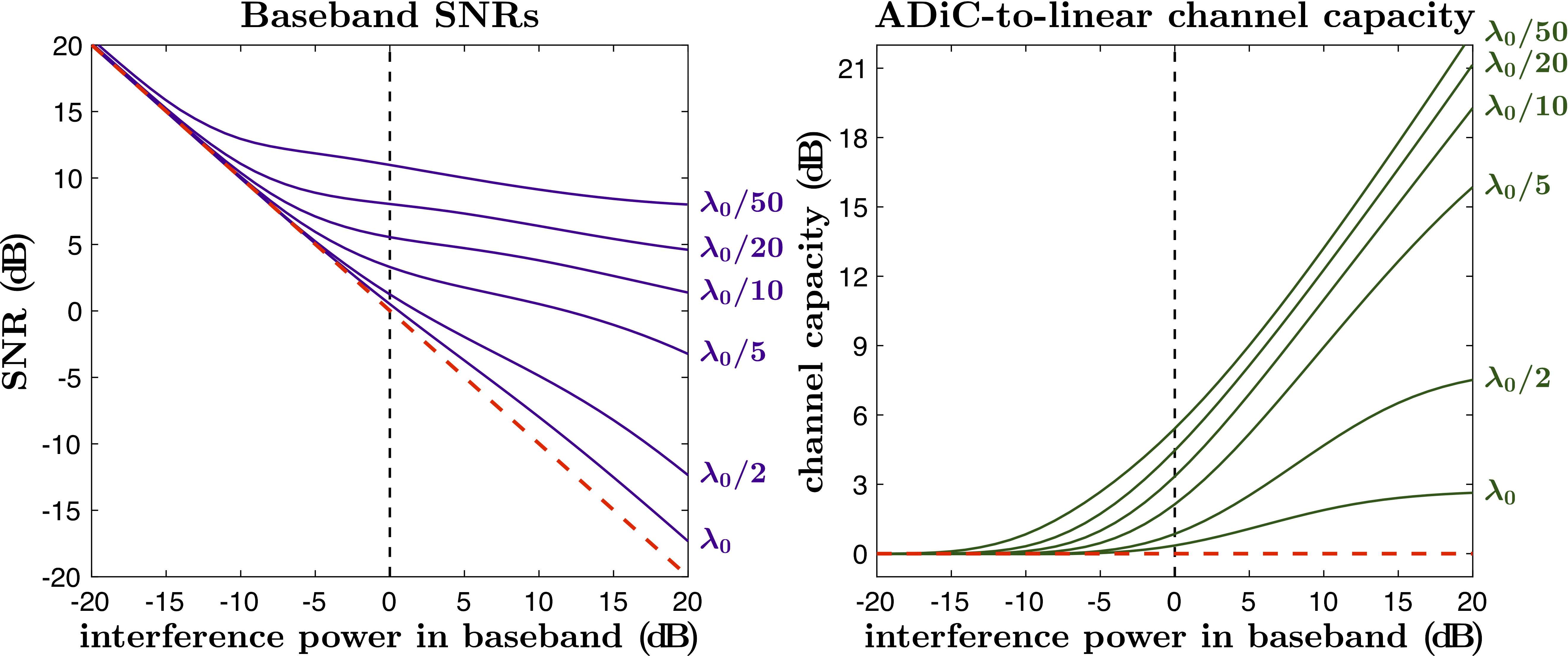}}
\caption{ADiC vs. linear SNRs and channel capacities for ``shared band".
\label{fig:shared band capacities}}
\end{figure}
\begin{figure*}[!t]
\centering{\includegraphics[width=14.6cm]{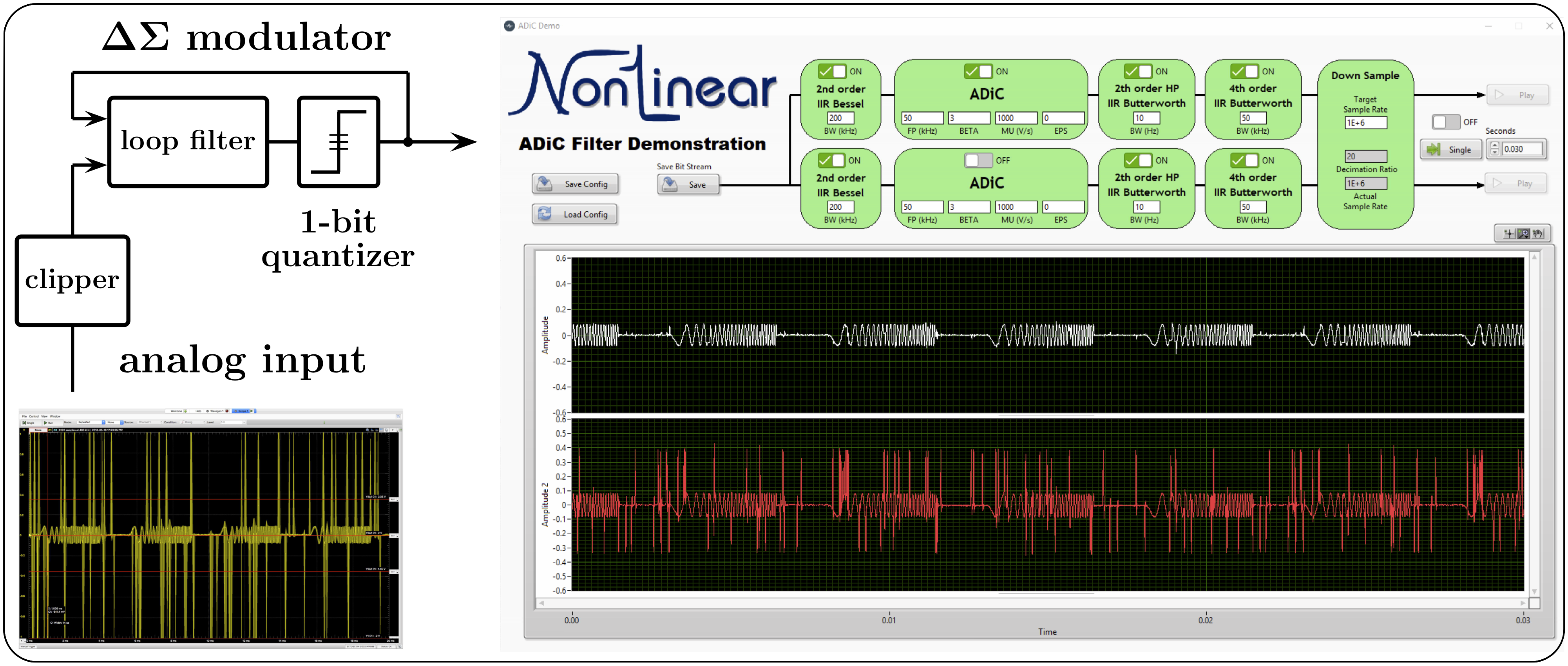}}
\caption{Prototype of ADiC filtering demo board.
\label{fig:demo}}
\end{figure*}
\begin{figure}[!b]
\centering{\includegraphics[width=8.6cm]{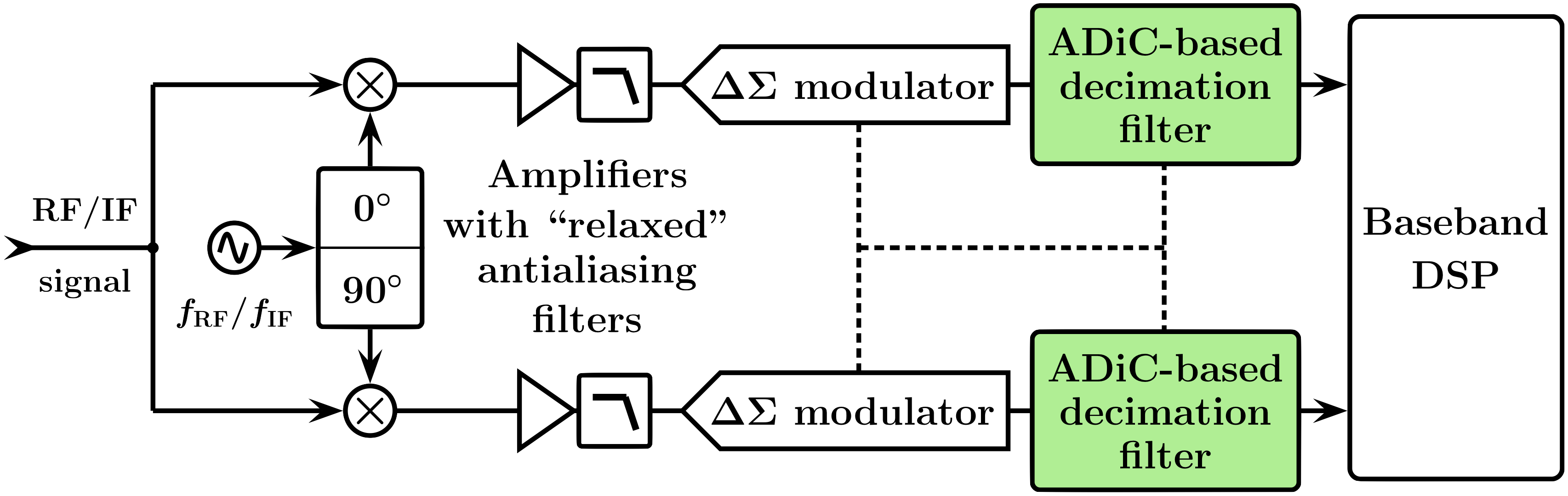}}
\caption{ADiC-based filtering in quadrature receiver.
\label{fig:quadrature receiver}}
\end{figure}

\subsection{``Shared band" case} \label{subsec:shared band}
While the main focus of the ADiC-based filtering is mitigation of wideband outlier noise affecting a band-limited signal of interest, this filtering can also be used to reduce outlier interference that, either intentionally or by system constraints, is confined to the signal's band. In such a case, for example, the signal+noise mixture can be treated simply as noise containing outliers, and an ADiC can be used to mitigate the outliers. As illustrated in Fig.~\ref{fig:shared band traces}, the baseband filter will have a negligible effect on such a mixture, while an ADiC deployed ahead of the baseband filter can suppress the narrow-band outliers and improve the SNR. Fig.~\ref{fig:shared band capacities} illustrates the improvements in the baseband SNRs and in the channel capacities, as functions of the interference power in the baseband, for a narrow-band Poisson impulsive noise with different rates. In this example $\lambda_0=2.27 B_0=\lambda_{\rm c}/10$, where $\lambda_{\rm c}$~is the ``pileup threshold rate" of the previous simulations, and the thermal noise is negligible so that the SNR is determined by the interference only. Note that, since suppression of the ``shared band" outliers requires that they are apparent in the baseband, both the mitigable event rates and the mitigable SNRs for narrow-band outliers are much lower than those for wideband outliers.

\subsection{Designing development \& testing platform} \label{subsec:prototype}
Fig.~\ref{fig:demo} shows an early prototype of an ADiC development and demonstration board that uses the ``effectively analog" implementation approach outlined in Section~\ref{subsubsec:bandwidth}. This board employs the 1-bit isolated 2nd~order $\Delta\Sigma$ modulator AD7403, implements ADiC-based filtering in FPGA using National Instruments' (NI's) reconfigurable I/O (RIO) controller board NI sbRIO-9637, programmed using NI's LabVIEW graphical development environment and LabVIEW FPGA module. This allows fast and easy reconfigurability of the ADiC-based processing for evaluating the performance of alternative ADiC topologies and their dependence on the ADiC parameters. In addition to testing and displaying the comparative results of the ADiC-based filtering for various waveforms and noise compositions, in the frequency range for up to several hundreds of kilohertz, this board allows real-time audible range demonstrations with instant playback. This development board is a step toward application-specific ADiC configurations, e.g. superheterodyne and/or direct conversion receiver architectures with quadrature baseband ADCs illustrated in Fig.~\ref{fig:quadrature receiver}. Since the power of transient interference is shared between the in-phase and the quadrature channels in the receiver, the complex-valued processing (as opposed to separate processing of the in-phase/quadrature components) has a potential for greatly improving the efficiency of the ADiC-based interference mitigation~\cite{Nikitin19ADiCpatentCIPs}. Such complex-valued processing is indicated by the dashed lines in Fig.~\ref{fig:quadrature receiver}.

\begin{figure}[!b]
\centering{\includegraphics[width=8.6cm]{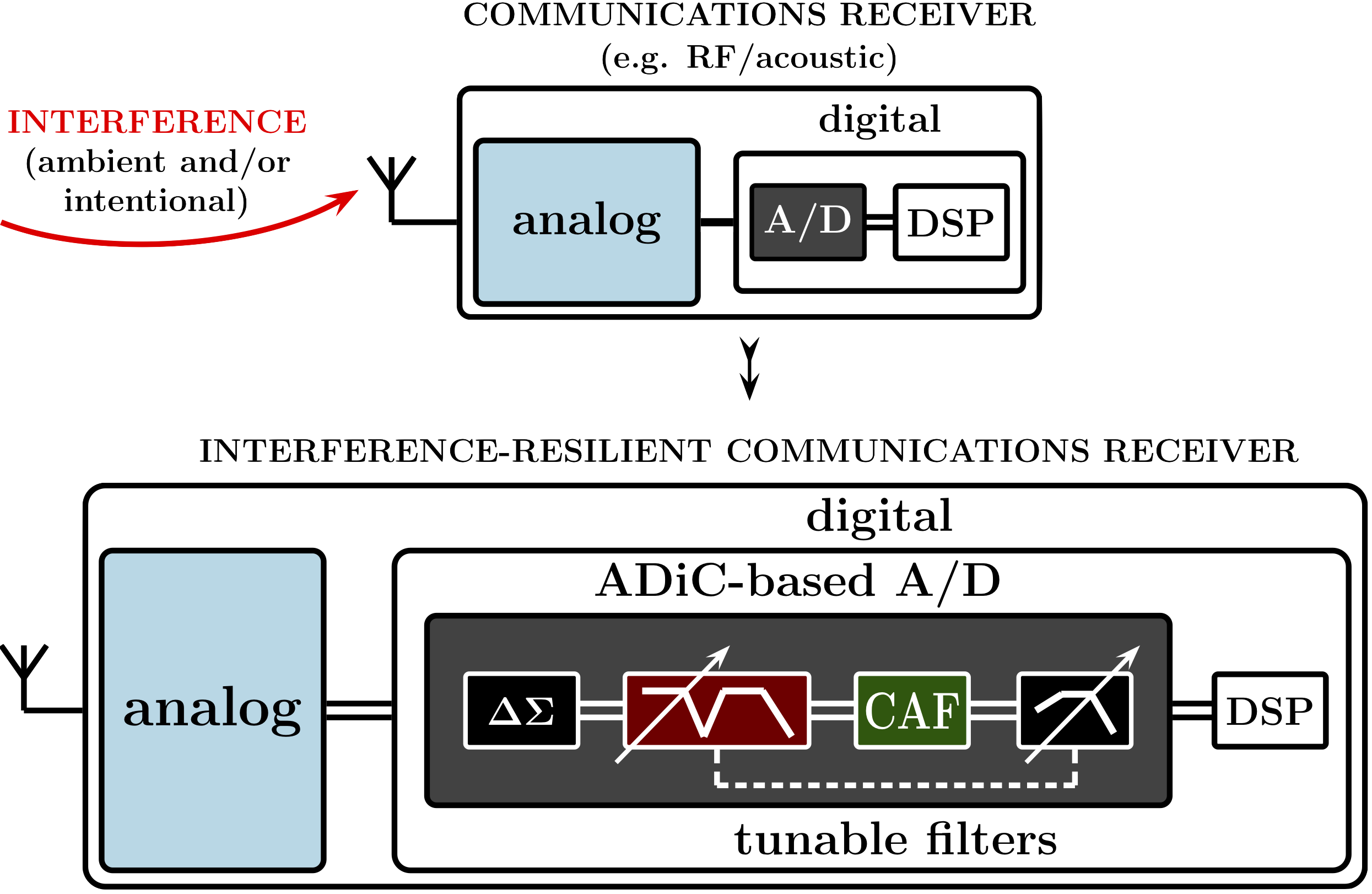}}
\caption{Interference resilient ADiC-based receiver.
\label{fig:InterferenceResilient}}
\end{figure}

Fig.~\ref{fig:InterferenceResilient} summarizes the potential use of the ADiC-based A/D conversion for development of communication receivers resilient to outlier interference of various types and origins, including those due to intermodulation distortions (IMD) and spectral regrowth caused by strong signals. This approach can be integrated into existing communication systems, e.g. implemented with existing communication radios operating in the HF, VHF and UHF spectrum. As discussed in Section~\ref{subsubsec:complex}, tunable wideband linear filters can be deployed ahead of the CAF for outlier enhancement, and various machine learning and optimization-based techniques can be used for their tuning to optimize the receiver performance for particular system configurations and/or interference scenarios.

\section{Conclusion} \label{sec:conclusion}
This paper provides an overview of the methodology and tools, including their analog and digital implementations, for real-time mitigation of outlier interference in general and ``hidden" wideband outlier noise in particular. Such mitigation is performed as a ``first line of defense" against interference ahead of, or in the process of, reducing the bandwidth to that of the signal of interest. Either used by itself, or in combination with subsequent interference mitigation techniques, this approach provides interference mitigation levels otherwise unattainable, with the effects, depending on particular interference scenarios, ranging from ``no harm" to considerable. While the main focus of this filtering technique is mitigation of wideband outlier noise affecting a band-limited signal of interest, it can also be used, given some {\it a priori} knowledge of the signal of interest's structure, to reduce outlier interference that is confined to the signal's band.

A distinct feature of the proposed approach is complementary filtering that capitalizes on the ``excess band" observation of wideband outlier noise for its efficient in-band mitigation by intermittently nonlinear filters. This significantly extends the mitigation range, in terms of both the rates of the outlier generating events and the mitigable SNRs, in comparison with the mitigation techniques focused on the apparent in-band effects of outlier interference. For example, the mitigable rates can be increased by more that an order of magnitude, and efficient mitigation can be performed for outlier noise in-band SNRs exceeding 30\,dB.

While the proposed filtering structures are mostly ``blind" and do not rely on any assumptions for the underlying noise distribution beyond its ``outlier" origins, they are adaptable to nonstationary signal and noise conditions and to various complex signal and interference mixtures. In particular, they can be used with the ``no harm" constraint such that nonlinear filtering does not degrade the resulting signal quality, as compared with the linear filtering, for a wide range of signal+noise compositions. This allows us to avoid the detrimental effects, such as distortions and instabilities, often associated with nonlinear filtering. The ``no harm" property is especially important when addressing widely and/or rapidly fluctuating interference conditions, e.g.  in mobile and cognitive communication systems.

The presented filters can be successfully used to suppress interference from diverse sources, including the RF co-site interference and the platform noise generated by on-board digital circuits, clocks, buses, and switching power supplies. They can also help to address multiple spectrum sharing and coexistence applications (e.g. radar-communications, radar-radar, narrowband/UWB, etc.), including those in dual function systems (e.g. when using radar and communications as mutual signals of opportunity). They can further benefit various other military, scientific, industrial, and consumer systems such as sensors/sensor networks and coherent imaging systems, sonar and underwater acoustic communications, auditory tactical communications, radiation detection, powerline communications, navigation and time-of-arrival techniques, and many others.

Finally, various embodiments of the presented ADiC-based filters allow relatively simple analog and/or real-time digital implementations. Thus they can be integrated into, and manufactured as IC components for use in different products, e.g. as A/D converters with incorporated interference suppression.

\section*{Acknowledgment}
The authors would like to thank
Keith~W. Cunningham of Atkinson Aeronautics \& Technology Inc., Fredericksburg, VA;
Scott~C. Geier of Pyvonics LLC, Garden City, KS;
James~E. Gilley of BK Technologies, West Melbourne, FL;
William~B. Kuhn of Kansas State University, Manhattan, KS;
Earl McCune of Eridan Communications, Santa Clara, CA;
Arlie Stonestreet\,\,II of Ultra Electronics ICE, Manhattan, KS,
and Kyle~D. Tidball of Mid-Continent Instruments and Avionics, Wichita, KS,
for their valuable suggestions and critical comments.

\small

\end{document}